\documentclass[11pt,draftclsnofoot,onecolumn]{IEEEtran}

\input{mysymbol.sty}
\usepackage{graphicx,slashbox}
\usepackage{epsfig}
\usepackage{amsmath}
\usepackage{bbm,dsfont}
\usepackage{amssymb}
\usepackage{subfigure}
\usepackage{wrapfig}
\usepackage{times,color}
\usepackage{cite,footnote,xspace,syntonly,algorithm,algorithmic,bm}
\usepackage{algorithm,algorithmic}
\usepackage[caption=false,font=footnotesize]{subfig}

\newcommand{\calN}{{\cal N}}

\def \cL {\mathcal{L}}

\def \cS {\mathcal{S}}

\def \cP {\mathcal{P}}

\def \bY {{\bf Y}}
\def \bR {{\bf R}}
\def \bX {{\bf X}}
\def \bA {{\bf A}}
\def \bV {{\bf V}}
\def \bW {{\bf W}}
\def \bF {{\bf F}}

\def \bI {{\bf I}}
\def \bB {{\bf B}}
\def \bM {{\bf M}}
\def \bU {{\bf U}}
\def \bV {{\bf V}}
\def \bE {{\bf E}}

\def \bZ {{\bf Z}}
\def \bW {{\bf W}}
\def \bH {{\bf H}}

\def \bQ {{\bf Q}}
\def \bC {{\bf C}}

\def \bP {{\bf P}}
\def \bGamma {{\bf \Gamma}}
\def \bw {{\bf w}}
\def \bu {{\bf u}}

\def \bx {{\bf x}}

\def \bb {{\bf b}}

\def \be {{\bf e}}

\def \bu {{\bf u}}

\def \bSigma {{\bf \Sigma}}

\def \ind {\mathbbm{1}}
\def \sgn {\text{sgn}}

\def \vec {\text{vec}}

\def \supp {\text{supp}}
\def \Phiper {\Phi^{\bot}}
\def \Omegaper {\Omega^{\bot}}

\long\def\symbolfootnote[#1]#2{\begingroup
\def\thefootnote{\fnsymbol{footnote}}
\footnote[#1]{#2}\endgroup} \psfull

\begin{document}
\title{\huge Recovery of Low-Rank Plus Compressed Sparse Matrices with Application
to Unveiling Traffic Anomalies$^\dag$}

\author{{\it Morteza Mardani, Gonzalo~Mateos, and Georgios~B.~Giannakis~(contact author)$^\ast$}}

\markboth{IEEE TRANSACTIONS ON INFORMATION THEORY (SUBMITTED)}
\maketitle \maketitle \symbolfootnote[0]{$\dag$ Work in this
paper was supported by the MURI Grant No. AFOSR FA9550-10-1-0567.
Parts of the paper 
will appear in the {\it IEEE Statistical Signal Processing Workshop}, Ann Arbor, MI,
Aug. 5-8, 2012.} \symbolfootnote[0]{$\ast$ The authors are with the Dept.
of ECE and the Digital Technology Center, University of
Minnesota, 200 Union Street SE, Minneapolis, MN 55455. Tel/fax:
(612)626-7781/625-4583; Emails:
\texttt{\{morteza,mate0058,georgios\}@umn.edu}}

\vspace*{-80pt}
\begin{center}
\small{\bf Submitted: }\today\\
\end{center}
\vspace*{10pt}


\thispagestyle{empty}\addtocounter{page}{-1}
\begin{abstract}
Given the superposition of a low-rank matrix plus the product of a
known fat compression matrix times a sparse matrix, the goal of this
paper is to establish deterministic conditions under which exact recovery of the
low-rank and sparse components becomes possible. This fundamental
identifiability issue arises with traffic anomaly detection in backbone networks,
and subsumes compressed sensing as well as the timely
low-rank plus sparse matrix recovery tasks encountered in matrix
decomposition problems. Leveraging the ability of $\ell_1$- and nuclear
norms to recover sparse and low-rank matrices, a convex program is
formulated to estimate the unknowns. Analysis and simulations confirm that the said
convex program can recover the unknowns for sufficiently low-rank and sparse enough components, along with a compression
matrix possessing an isometry property when restricted to operate on sparse
vectors. When the low-rank, sparse, and compression matrices are
drawn from certain random ensembles, it is established that exact recovery is possible
with high probability. First-order algorithms are developed to solve the nonsmooth
convex optimization problem with provable iteration complexity guarantees.
Insightful tests with synthetic and real network data corroborate
the effectiveness of the novel approach in unveiling traffic anomalies across flows and time,
and its ability to outperform existing alternatives.
\end{abstract}
\vspace*{-5pt}
%
\newpage


\section{Introduction}
\label{sec:intro}

Let $\bX_0 \in \mathbb{R}^{L \times T}$ be a low-rank matrix
[$r:={\rm rank}(\bX_0) \ll \min(L,T)$], and let $\bA_0 \in \mathbb{R}^{F \times T}$
be sparse ($s:=\|\bA_0\|_0 \ll FT$, $\|\cdot\|_0$ counts the
nonzero entries of its matrix argument). Given a compression
matrix $\bR \in \mathbb{R}^{L \times F}$ with $L\leq F$, and observations
\begin{align}
\bY=\bX_0+\bR\bA_0  \label{eq:model}
\end{align}
the present paper deals with the recovery of $\{\bX_0,\bA_0\}$. This task is of interest e.g., to unveil
anomalous flows in backbone networks~\cite{LCD04,MMG11,zggr05}, to
extract the time-varying foreground from a sequence of compressed video
frames~\cite{Branuick_nips11},
or, to identify active brain regions from undersampled functional
magnetic resonance imagery (fMRI)~\cite{Vaswani_Allerton_11}.
In addition, this fundamental problem is found at the crossroads of
compressive sampling (CS), and the timely low-rank-plus-sparse matrix decompositions.

In the absence of the low-rank
component ($\bX_0=\mathbf{0}_{L\times T}$), one is left with an
under-determined sparse signal recovery problem; see e.g.,~\cite{CT05,rauhut}
and the tutorial account~\cite{candes_tutorial}.
When $\bY=\bX_0+\bA_0$, the formulation boils down to principal
components pursuit (PCP), also referred to as robust principal component
analysis (PCA)~\cite{CLMW09,CSPW11,bayes_rpca}. For this idealized
noise-free setting, sufficient conditions for
exact recovery are available for both of the aforementioned special
cases. However, the superposition of a low-rank and a
\textit{compressed} sparse matrix in \eqref{eq:model} further challenges identifiability
of $\{\bX_0,\bA_0\}$. In the presence of `dense' noise,
stable reconstruction of the low-rank and sparse matrix components
is possible via PCP~\cite{zlwcm10,Outlier_pursuit}. Earlier efforts dealing with
the recovery of sparse vectors in noise led to similar performance
guarantees; see e.g.,~\cite{bickel09} and references therein.
Even when $\bX_0$ is nonzero, one could envision a CS variant where the
measurements are corrupted with correlated (low-rank)
noise~\cite{Vaswani_Allerton_11}. Last but not least, when $\bA_0=\mathbf{0}_{F\times T}$
and $\bY$ is noisy, the recovery of $\bX_0$ subject to a rank
constraint is nothing else than PCA -- arguably, the
workhorse of high-dimensional data analysis~\cite{J02}.

The main contribution of this paper is to establish that given
$\bY$ and $\bR$ in \eqref{eq:model},
for small enough $r$ and $s$ one can \textit{exactly}
recover $\{\bX_{0},\bA_0\}$ by solving the nonsmooth \textit{convex} optimization
problem
\begin{align}
\mathrm{(P1)} ~~~~ \min_{\{\bX,\bA\}}&~~~ \|\bX\|_{\ast}+\lambda \|\bA\|_1\nonumber\\
\mathrm{s. to} &~~~ \bY=\bX+\bR \bA  \nonumber
\end{align}
where $\lambda \geq 0$ is a tuning parameter; $\|\bX\|_{\ast}:=\sum_{i}\sigma_i(\bX)$ is
the nuclear norm of $\bX$ ($\sigma_i$  stands for the $i$-th
singular value); and, $\|\bX\|_1:=\sum_{i,j}|x_{ij}|$ denotes the
$\ell_1$-norm. The aforementioned norms are convex surrogates to the rank and
$\ell_0$-norm, respectively, which albeit natural as criteria they are
NP-hard to optimize~\cite{l_0_NP_hard,rank_NP_Duro}. Recently, a
greedy algorithm for recovering low-rank and sparse matrices from
compressive measurements was put forth in~\cite{Branuick_nips11}.
However, convergence of the algorithm and its error performance are
only assessed via numerical simulations. A recursive algorithm capable
of processing data in real time can be found in~\cite{Vaswani_Allerton_11},
which attains good performance in practice but does not offer theoretical
guarantees.

A {\it deterministic} approach along the lines of~\cite{CSPW11} is
adopted first to derive conditions under which \eqref{eq:model} is
locally identifiable (Section \ref{sec:local_id}). Introducing
a notion of incoherence between the additive components $\bX_0$ and
$\bR\bA_0$, and resorting to the restricted isometry constants of
$\bR$~\cite{CT05}, sufficient conditions are obtained to ensure that (P1) succeeds in exactly
recovering the unknowns (Section
\ref{subsec:mainresult}). Intuitively, the results here assert that
if $r$ and $s$ are sufficiently small, the nonzero entries of
$\bA_0$ are sufficiently spread out, and subsets of columns of $\bR$
behave as isometries, then (P1) exactly recovers $\{\bX_0,\bA_0\}$. As
a byproduct, recovery results for PCP and CS are also obtained
by specializing the aforesaid conditions accordingly (Section \ref{subsec:induced_results}).
The proof of the main result builds on Lagrangian duality theory~\cite{Bers,Boyd},
to first derive conditions under which $\{\bX_0,\bA_0\}$ is the \emph{unique}
optimal solution of (P1) (Section \ref{ssec:proof_unique_optimality}).
In a nutshell, satisfaction of the optimality conditions
is tantamount to the existence of a valid dual certificate.
Stemming from the unique challenges introduced by $\bR$, the dual certificate construction
procedure of Section \ref{ssec:proof_dualcert_const} is markedly distinct from the direct sum approach in~\cite{CSPW11}, and
the (random) golfing scheme of~\cite{CLMW09}.
Section \ref{sec:sat} shows that low-rank, sparse, and compression matrices
drawn from certain random ensembles satisfy the sufficient conditions
for exact recovery with high probability.

Two iterative algorithms for solving (P1) are developed in Section \ref{sec:alg},
which are based on the accelerated proximal grandient (APG)
method~\cite{nesterov83,nesterov05,fista,rpca_proximal}, and
the alternating-direction method of multipliers (AD-MoM)~\cite{Bertsekas_Book_Distr,Boyd}.
Numerical tests corroborate the exact recovery claims, and the effectiveness
of (P1) in unveiling traffic volume anomalies from real network data (Section \ref{sec:sims}).
Section \ref{sec:disscusion} concludes the paper with a summary and
a discussion of limitations, possible extensions, and interesting future directions.
Technical details are deferred to the Appendix.


\subsection{Notational conventions}
\label{ssec:notation}

Bold uppercase (lowercase) letters will denote matrices (column vectors), and
calligraphic letters will denote sets.
Operators $(\cdot)'$, $(\cdot)^\dagger$, $\textrm{tr}(\cdot)$, $\textrm{vec}(\cdot)$,
$\rm{diag(\cdot)}$, $\lambda_{\rm max}(\cdot)$, $\sigma_{\rm min}(\cdot)$, and $\otimes$
will denote transposition, matrix pseudo inverse, matrix trace, matrix vectorization,
diagonal matrix, spectral radius, minimum singular value, and Kronecker
product, respectively; $|\cdot|$ will be used for the cardinality of a set and
the magnitude of a scalar. The $n \times n$ identity matrix will be represented by $\bI_n$
and its $i$-th column by $\be_i$; while $\mathbf{0}_{n}$ denotes the
$n \times 1$ vector of all zeros, and $\mathbf{0}_{n \times p}:=\mathbf{0}_{n} \mathbf{0}'_{p}$.
The $\ell_q$-norm of vector $\mathbf{x}\in\mathbb{R}^{p}$ is
$\|\mathbf{x}\|_q:=\left(\sum_{i=1}^p|x_i|^q\right)^{1/q}$ for
$q\geq 1$.
For matrices $\bA,\bB \in \mathbb{R}^{m \times n}$
define the trace inner product $\langle \bA,\bB \rangle:=\mbox{tr}(\bA' \bB)$.
Also, recall that $\|\bA\|_F:=\sqrt{\mbox{tr}\left(\bA\bA'\right)}$ is the Frobenious norm,
$\|\bA\|_1:=\sum_{i,j} |a_{ij}|$
is the $\ell_1$-norm, $\|\bA\|_{\infty}:=\max_{i,j} |a_{ij}|$ is the $\ell_{\infty}$-norm, and
$\|\bA\|_{\ast}:=\sum_{i}\sigma_i(\bA)$ is the nuclear norm. In addition,
$\|\bA\|_{1,1}:=\max_{\|\bx\|_1=1} \|\bA\bx\|_1=\max_i\|\be_i'\bA \|_1$
denotes the induced $\ell_1$-norm, and
likewise for the induced $\ell_{\infty}$-norm
$\|\bA\|_{\infty,\infty}:=\max_{\|\bx\|_{\infty}=1} \|\bA\bx\|_{\infty}=\max_i\|\bA \be_i\|_1$.
For the linear operator $\mathcal{A}$, define the operator norm
$\|\mathcal{A}\|:=\max_{\|\bX\|_F=1}\|\mathcal{A}(\bX)\|_F$, which subsumes
the spectral norm $\|\bA\|:=\max_{\|\bx\|=1} \|\bA\bx\|$.
Define also the support set $\supp(\bA):=\{(i,j):a_{ij} \neq 0\}$. The indicator
function $\ind_{\{a=b\}}$ equals one when $a=b$, and zero otherwise.


\section{Local Identifiability}
\label{sec:local_id}

The first issue to address is model identifiability, meaning that there are
\emph{unique} low-rank and sparse matrices satisfying
\eqref{eq:model}. If there exist multiple decompositions of
$\bY$ into $\bX+\bR \bA$ with low-rank $\bX$ and sparse $\bA$, there is no hope
of recovering $\{\bX_0,\bA_0\}$ from the data. For instance, if the null space of
the fat matrix $\bR$ contains sparse matrices, there may exist a sparse perturbation $\bH$
such that $\bA_0+\bH$ is still sparse and $\{\bX_{0},\bA_0+\bH\}$
is a legitimate solution. Another problematic case
arises when there is a sparse perturbation $\bH$ such that $\bR\bH$ is spanned by the
row or column spaces of $\bX_0$. Then, $\bX_0+\bR\bH$ has the same rank as
$\bX_0$ and $\bA_0-\bH$ may still be sparse. As a result, one may pick
$\{\bX_0+\bR\bH,\bA_0-\bH\}$ as another valid solution.
Dealing with such identifiability issues is the subject of this section.

Let $\bU\bSigma\bV'$ denote the singular value decomposition (SVD)
of $\bX_0$, and consider the subspaces: s1)
$\Phi(\bX_0) := \{\bZ\in\mathbb{R}^{L\times T}:\bZ=\bU \bW_1' +
\bW_2 {\bV}',\:\bW_1 \in \mathbb{R}^{T \times r},\:\bW_2 \in
{\mathbb{R}}^{L \times r}\}$ of matrices in
either the column or row space of $\bX_0$; s2) $\Omega(\bA_0)
:=\{\bH\in\mathbb{R}^{F\times T}:\supp(\bH) \subseteq \supp(\bA_0)\}$ of
matrices in $\mathbb{R}^{F\times T}$
with support contained in the support of $\bA_0$;
and s3) $\Omega_R(\bA_0) := \{\bZ\in\mathbb{R}^{L\times T}:\bZ=\bR \bH,
\:\bH \in \Omega(\bA_0)\}$. For notational brevity, s1)-s3)
will be henceforth denoted as $\{\Phi,\Omega,\Omega_R\}$.
Noteworthy properties of these subspaces are:
i) both $\Phi$ and $\Omega_R \subset \mathbb{R}^{L\times T}$,
hence it is possible to directly compare elements from them;
ii) $\bX_0 \in \Phi$ and $\bR \bA_0 \in \Omega_R$; and iii)
if $\bZ\in\Phiper$ is added to $\bX_0$,
then $\textrm{rank}(\bZ+\bX_0)>r$.

For now, assume that the subspaces $\Omega_R$ and $\Phi$ are also
known. This extra information helps identifiability of
\eqref{eq:model}, because potentially troublesome solutions
$\{\bX_0+\bR\bH,\bA_0-\bH\}$ are limited to a restricted class. If
$\bX_0+\bR\bH \notin \Phi$ or $\bA_0-\bH \notin \Omega$, that
candidate solution is not admissible since it is known a priori that
$\bA_0 \in \Omega$ and $\bX_0 \in \Phi$. Under these assumptions,
the following lemma puts forth the necessary and sufficient conditions
guaranteeing unique decomposability of $\bY$ according to
\eqref{eq:model} -- a notion known as {\it local
identifiability}~\cite{CLMW09}.

\begin{lemma}\label{lem:lem_1}
Matrix $\bY$ uniquely decomposes into $\bX_0+\bR \bA_0$ if and
only if $\Phi \cap \Omega_R=\{\mathbf{0}_{L \times T}\}$,
and ${\bf{R}\bf{H}} \neq \mathbf{0}_{L \times T},
\forall {\bf{H}} \in \Omega \backslash\{\mathbf{0}_{F \times T}\}$.
\end{lemma}

\begin{IEEEproof}
Since by definition $\bX_0 \in \Phi$ and $\bA_0 \in \Omega$,
one can represent every element in the \emph{subspaces} $\Phi$ and
$\Omega_R$ as $\bX_0+\bZ_1$ and $\bR\bA_0+\bZ_2$, respectively, where $\bZ_1 \in \Phi$
and $\bZ_2 \in \Omega_R$. Assume that $\Phi \cap \Omega_R=\{\mathbf{0}_{L \times T}\}$,
and suppose by contradiction that there exist {\it nonzero} perturbations
$\{\bZ_1,\bZ_2\}$ such that $\bY=\bX_0+\bZ_1+\bR\bA_0+\bZ_2$. Then,
$\bZ_1+\bZ_2=\mathbf{0}_{L\times T}$, meaning that $\bZ_1$ and $\bZ_2$
belong to the same subspace, which contradicts the assumption.
Conversely, suppose there exists a non-zero $\bZ \in \Omega_R \cap \Phi$.
Clearly, $\{\bX_0+\bZ,\bR\bA_0-\bZ\}$ is a feasible solution
where $\bX_0+\bZ \in \Phi$ and $\bR\bA_0-\bZ \in \Omega_R$. This contradicts the
uniqueness assumption. In addition, the condition
${\bf{R}\bf{H}} \neq \mathbf{0}, {\bf{H}} \in \Omega\backslash\{\mathbf{0}_{L\times T}\}$
ensures that $\bZ=\mathbf{0}_{L\times T} \in \Phi \cap \Omega_R$ only when
$\bZ=\bR\bH=\mathbf{0}_{L\times T}$ for $\bH=\mathbf{0}_{F \times T}$.
\end{IEEEproof}

In words, \eqref{eq:model} is locally identifiable if and only
if the subspaces $\Phi$ and $\Omega_R$ intersect transversally, and the
sparse matrices in $\Omega$ are not annihilated by $\bR$. This last condition
is unique to the setting here, and is not present in~\cite{CLMW09} or~\cite{CSPW11}.

\begin{remark}[Projection operators]\label{remark:proj}
\normalfont Operator $\cP_{\Omega}(\bX)$ ($\cP_{\Omegaper}(\bX)$) denotes
the orthogonal projection of $\bX$ onto the subspace $\Omega$ (orthogonal
complement $\Omegaper$). It simply sets those elements of $\bX$
not in $\textrm{supp}(\bA_0)$ to zero. Likewise,
$\cP_{\Phi}(\bX)$ ($\cP_{\Phiper}(\bX)$) denotes the orthogonal projection
of $\bX$ onto the subspace $\Phi$ (orthogonal complement $\Phiper$).
Let $\bP_U:=\bU\bU'$ and $\bP_V:=\bV\bV'$ denote, respectively,
projection onto the column and row spaces of $\bX_{0}$.
It can be shown that
$\cP_{\Phi}(\bX)=\bP_U\bX+\bX\bP_V-\bP_U\bX\bP_V$, while
the projection onto the complement subspace
is $\cP_{\Phiper}(\bX)=(\bI-\bP_U)\bX(\bI-\bP_V)$. In addition,
the following identities
\begin{align}
\langle \cP_{\Phi}(\bX), \cP_{\Phi}(\bY) \rangle = \langle \cP_{\Phi}(\bX), \bY \rangle = \langle \bX, \cP_{\Phi}(\bY)
\rangle \label{eq:proj_exchange}
\end{align}
of orthogonal projection operators such as $\cP_{\Phi}(\cdot)$,
will be invoked throughout the paper.
\end{remark}


\subsection{Incoherence measures}
\label{ssec:coh}

Building on Lemma~\ref{lem:lem_1}, alternative sufficient conditions are
derived here to ensure local identifiability.
To quantify the overlap between $\Phi$ and $\Omega_R$,
consider the {\it incoherence} parameter
\begin{equation}
 \mu (\Omega_R,\Phi)=\max_{\bZ \in {\Omega_R \backslash \{\mathbf{0}\}} }
 \frac{\|\cP_{\Phi}(\bZ)\|_F}{\|\bZ\|_F}. \label{eq:mau_mat}
\end{equation}
for which it holds that $\mu (\Omega_R,\Phi) \in [0,1]$. The lower bound is
achieved when $\Phi$ and $\Omega_R$ are orthogonal, while the upper
bound is attained when $\Phi \cap \Omega_R$ contains a nonzero
element. Assuming $\Phi \cap \Omega_R = \{\mathbf{0}_{L\times T}\}$,
then $\mu (\Omega_R,\Phi) < 1$ represents the cosine of the angle
between $\Phi$ and $\Omega_R$~\cite{Deutsch}. From Lemma
\ref{lem:lem_1}, it appears that $\mu (\Omega_R,\Phi) < 1$
guarantees $\Phi \cap \Omega_R = \{\mathbf{0}_{L\times T}\}$. As it
will become clear later on, tighter conditions on $\mu
(\Omega_R,\Phi)$ will prove instrumental to guarantee exact recovery
of $\{\bX_0,\bA_0\}$ by solving (P1).

To measure the incoherence among subsets of columns of
$\bR$, which is tightly related to the
second condition in Lemma~\ref{lem:lem_1}, the restricted
isometry constants (RICs) come handy~\cite{CT05}. The constant
$\delta_k(\bR)$ measures the extent to which a $k$-subset of
columns of $\bR$ behaves like an isometry. It is defined
as the smallest value satisfying
\begin{equation}
c(1-\delta_k(\bR)) \leq \frac{\|\bR \bu\|^2}{\|\bu\|^2} \leq c(1+\delta_k(\bR)) \label{eq:rip_ineq}
\end{equation}
for every $\bu \in \mathbb{R}^F$ with $\|\bu\|_0 \leq k$ and for
some positive normalization constant $c<1$~\cite{CT05}. For later
use, introduce $\theta_{s_1,s_2}(\bR)$ which measures `how
orthogonal' are the subspaces generated by two disjoint column subsets of $\bR$, with
cardinality $s_1$ and $s_2$. Formally, $\theta_{s_1,s_2}(\bR)$ is the smallest value that satisfies
\begin{align}
|\langle \bR \bu_1, \bR \bu_2 \rangle| \leq c \theta_{s_1,s_2}(\bR) \|\bu_1\| \|\bu_2\| \label{eq:theta}
\end{align}
for every $\bu_1,\bu_2 \in \mathbb{R}^F$, where $\supp(\bu_1) \cap
\supp(\bu_2)=\emptyset$ and $\|\bu_1\|_0 \leq s_1,\|\bu_2\|_0 \leq
s_2$. The normalization constant $c$ plays the same role as in
$\delta_k(\bR)$. A wide family of matrices with small RICs have been
introduced in e.g.,~\cite{CT05}.

All the elements are now in place to state this section's main result.

\begin{proposition}\label{prop:prop_1}
Assume that each column of $\bA_0$ contains at most $k$ nonzero elements.
If $\mu(\Omega_R,\Phi) < 1$ and $\delta_k(\bR)  <1$, then
$\Omega_R \cap \Phi = \{\mathbf{0}_{L\times T}\}$ and
${\bf{R}\bf{H}} \neq {\mathbf{0}_{L\times T}}, {\bf{H}}
\in \Omega \backslash\{{\mathbf{0}}_{F \times T}\}$.
\end{proposition}
\begin{IEEEproof} Suppose the intersection in nontrivial,
meaning that there exists nonzero matrices $\bH \in \Omega$
and $\bU \bW_1' + \bW_2 \bV' \in \Phi$ satisfying $\bR\bH=\bU \bW_1' + \bW_2 \bV'$.
Vectorizing the last equation and relying on the identity
$\vec(\bA\bX\bB)=(\bB' \otimes \bA) \vec(\bX)$, one obtains a
linear system of equations
\begin{equation}
[\bI_T \otimes \bR \: -\bI_T \otimes \bU \: -\bV \otimes \bI_L] \bw=\mathbf{0}_{LT} \label{eq:vect}
\end{equation}
where $\bw := [{\vec(\bH)}'~\vec(\bW_1') ~\vec(\bW_2')]'$.
Define an $LT \times FT$ matrix $\bC_1 := \bI_T \otimes \bR$ and the $LT \times (L+T)r$
matrix $\bC_2 := [-\bI_T \otimes \bU \: -\bV \otimes \bI_L]$.
The corresponding coefficients are $\bw_1 := \vec(\bH)$ and
$\bw_2 :=  [{\vec(\bW_1')}'\:\vec(\bW_2')]'$. Then, \eqref{eq:vect} implies
 there exists a $\bw_1 \neq \mathbf{0}_{FT}$ such that
$\bC_1 \bw_1 +\bC_2 \bw_2=\mathbf{0}_{LT}$.

Consider two cases:
i) $\bw_2=\mathbf{0}_{r(L+T)}$, and ii) $\bw_2 \neq \mathbf{0}_{r(L+T)}$.
Under i) $\bC_1 \bw_1=\mathbf{0}_{LT}$, and thus $\bR \bw_1^{(i)}=\mathbf{0}$
for some nonzero $\bw_1^{(i)}$ with $i\in \{1,2,...,T\}$ where
$\bw_1=[\bw_1^{(1)}...\bw_1^{(T)}]$. Therefore, if $\|\bw_1^{(i)}\|_0 \leq k$,
$\delta_k(\bR)<1$ implies that $\bw_1^{(i)}=\mathbf{0}_{LT}$,
which is a contradiction. For ii) $\mu(\Omega_R,\Phi)<1$ implies
that there is no $\bw_1$ with $\supp(\bw_1)\subseteq \supp(\mathrm{vec(\bA_0)})$
and $\bw_2 \in \mathbb{R}^{(L+T)r}$ such that $\bC_1\bw_1+\bC_2\bw_2=\mathbf{0}_{FT}$,
since otherwise $|\langle \bC_1\bw_1,\bC_2\bw_2 \rangle|=\|\bC_1\bw_1\| \|\bC_2\bw_2\|$
which leads to $\mu(\Omega_R,\Phi)=1$.
\end{IEEEproof}


\section{Exact Recovery via Convex Optimization}
\label{sec:exact}

In addition to $\mu (\Omega_R,\Phi)$,
there are other incoherence measures which play an important role in
the conditions for exact recovery. Consider a feasible solution
$\{\bX_0+a_{ij}\bR\be_i\be_j',\bA_0-a_{ij}\be_i\be_j'\}$, where
$(i,j) \notin \supp(\bA_0)$ and thus $a_{ij}\be_i\be_j' \notin
\Omega$. It may then happen that $a_{ij}\bR\be_i\be_j' \in \Phi$ and
${\rm rank}(\bX_0+a_{ij}\bR\be_i\be_j') = {\rm rank}(\bX_0)-1$,
while $\|\bA_0-a_{ij}\be_i\be_j'\|_0 = \|\bA_0\|_0 + 1$, challenging
identifiability when $\Phi$ and $\Omega_R$ are unknown.
Similar complications will arise if $\bX_0$ has a sparse row space that
could be confused with the row space of $\bA_0$. These issues motivate defining
\begin{align}
\gamma_{R}(\bU):=\max_{i,j}
\frac{\|\bP_{U}\bR\be_i\be_j'\|_F}{\|\bR\be_i\be_j'\|_F},\quad
\gamma(\bV):=\max_{i} \|\bP_{V}\be_i\|_F \nonumber
\end{align}
where $\gamma_R(\bU),\gamma(\bV) \leq 1$. The maximum of
$\gamma_R(\bU)$ $[\gamma(\bV)]$ is attained when
$\bR\be_i\be_j'$ $[\be_i]$ is in the column [row] space of $\bX_0$
for some $(i,j)$. Small values of $\gamma_R(\bU)$ and $\gamma(\bV)$
imply that the column and row spaces of $\bX_0$ do not contain the
columns of $\bR$ and sparse vectors, respectively.

Another identifiability issue arises when $\bX_0=\bR\bH$ for some
sparse matrix $\bH\in\Omega$. In this case, each column of $\bX_0$
is spanned by a few columns of $\bR$. Consider the parameter
\begin{align}
\xi_R(\bU,\bV) := \|\bR'\bU\bV'\|_{\infty} =
\max_{i,j}|{\be_i}'\bR'\bU\bV\be_j|. \nonumber
\end{align}
A small value of $\xi_R(\bU,\bV)$ implies that each column of
$\bX_0$ is spanned by sufficiently many columns of $\bR$. To
understand this property, suppose for simplicity that all nonzero
singular values of $\bX_0$ are identical and equal to $\sigma$,
say. The $k$-th column of $\bX_0$ is then $\sum_{i=1}^{r} \sigma
\bu_i v_{i,k}$, and its projection onto the $l$-th column of $\bR$
is
\begin{align}
\Big|\langle \bR\be_l, \sum_{i=1}^{r} \sigma \bu_i v_{i,k}
\rangle\Big| = \sigma \Big| \sum_{i=1}^{r} \langle \bR\be_l,
\bu_i \rangle v_{i,k}\Big| \leq \sigma \xi_R(\bU,\bV).\nonumber
\end{align}
Since the energy of $\sum_{i=1}^{r} \sigma \bu_i v_{i,k}$ is somehow
allocated along the directions $\bR\be_l$, if all the aforementioned
projections can be made arbitrarily small, then sufficiently many
nonzero terms in the expansion are needed to account for all this
energy.


\subsection{Main result}
\label{subsec:mainresult}

\begin{theorem}\label{th:th_1}
Consider given matrices $\bY\in\mathbb{R}^{L \times T}$ and
$\bR\in\mathbb{R}^{L \times F}$ obeying
$\bY=\bX_{0}+\bR\bA_0=\bU\bSigma\bV'+\bR\bA_0$, with
$r:=\text{rank}(\bX_{0})$ and $s:=\|\bA_0\|_0$. Assume that every
row and column of $\bA_0$ has at most $k$ nonzero elements, and that
$\bR$ has orthonormal rows. If the following conditions
\begin{description}
\item[I)] $(1-\mu(\Phi,\Omega_R))^2 (1-\delta_k(\bR)) > \omega_{\text{max}}$; and
\item[II)] $(1+\alpha_{\text{max}})\left(\frac{1+\beta_{\text{max}}}{1-\beta_{\text{max}}}\right)
\xi_R(\bU,\bV) \sqrt{s} + \mu(\Phi,\Omega_R) (1+\delta_k(\bR))^{1/2} (1+\alpha_{\text{max}}) \sqrt{r} < 1$
\end{description}
hold, where
\begin{align}
\omega_{\text{max}}&:=\theta_{1,1}(\bR) [\sqrt{2}k+s\gamma^2(\bV)] + (1+\delta_1(\bR))
\left[ \sqrt{2}k  \gamma_R^2(\bU) + k \gamma^2(\bV) + s \gamma_R^2(\bU)\gamma^2(\bV)  \right] \nonumber\\
\alpha_{\text{max}}&:= \left[\frac{1}{c(1-\delta_k(\bR))(1-\mu(\Phi,\Omega_R))^2}-1\right]^{1/2}
\nonumber\\
\beta_{\text{max}}&:= \frac{1}{(1-\mu(\Omega_R,\Phi))^2(1-\delta_k(\bR)) {\omega_{\text{max}}^{-1} - 1 }}    \nonumber
\end{align}
then there exists $\lambda>0$ for which
the convex program (P1) exactly recovers $\{\bX_0,\bA_0\}$.
\end{theorem}

Note that I) alone is already more stringent than the pair of conditions
$\mu(\Omega_R,\Phi) < 1$ and $\delta_k(\bR) <1$ needed
for local identifiability (cf. Proposition \ref{prop:prop_1}).
Satisfaction of the conditions in Theorem~\ref{th:th_1} hinges upon
the values of the incoherence parameters $\mu
(\Omega_R,\Phi),\gamma_R(\bU),\gamma(\bV),$ $\xi_R(\bU,\bV)$, and
the RICs $\delta_k(\bR)$ and $\theta_{1,1}(\bR)$. In particular,
$\{\omega_{\text{max}},\alpha_{\text{max}},\beta_{\text{max}}\}$ are
increasing functions of these parameters, and it is readily
observed from I) and II) that the smaller
$\{\omega_{\text{max}},\alpha_{\text{max}},\beta_{\text{max}}\}$
are, the more likely the conditions are met. Furthermore, the incoherence parameters are
increasing functions of the rank $r$ and sparsity level $s$. The RIC
$\delta_k(\bR)$ is also an increasing function of $k$, the maximum
number of nonzero elements per row/column of $\bA_0$. Therefore, for
sufficiently small values of $\{r,s,k\}$, the sufficient conditions
of Theorem~\ref{th:th_1} can be indeed satisfied.

It is worth noting that not only $s$, but also the position of the
nonzero entries in $\bA_0$ plays an important role in satisfying
I) and II). This is manifested through $k$, for which a small value
indicates the entries of $\bA_0$ are sufficiently spread out, i.e.,
most entries do not cluster along a few rows or columns of $\bA_0$.
Moreover, no restriction is placed on the magnitude of these
entries, since as seen later on it is only the positions that affect
optimal recovery via (P1).

\begin{remark}[Row orthonormality of $\bR$]\label{rem:ortho_R}
\normalfont Assuming $\bR\bR'=\bI_L$ is equivalent to supposing that $\bR$ is
full-rank. This is because for a full row-rank
$\bR=\bU_{R}\bSigma_{R}{\bV_R}'$, one can pre-multiply both sides
of~\eqref{eq:model} with $\bSigma_{R}^{-1}{\bU_{R}}'$ to obtain
$\tilde\bR:={\bV_R}'$ with orthonormal rows.
\end{remark}


\subsection{Induced recovery results for principal components pursuit and compressed sensing}
\label{subsec:induced_results}

Before delving into the proof of the main result, it is instructive
to examine how the sufficient conditions in Theorem~\ref{th:th_1}
simplify for the subsumed PCP and CS problems. In PCP
one has $\bR=\bI_L$, which implies $\Omega_R=\Omega$ and
$\delta_k(\bR)=\theta_{1,1}(\bR)=0$. To obtain sufficient conditions expressed only in terms
of $\mu(\Phi,\Omega)$, one can borrow the coherence conditions of~\cite{CLMW09} and readily arrive at the following result.

\begin{corollary}\label{corr:corr_1}
Consider given $\bY\in\mathbb{R}^{L \times T}$ obeying
$\bY=\bX_{0}+\bA_0=\bU\bSigma\bV'+\bA_0$, with
$r:=\text{rank}(\bX_{0})$ and $s:=\|\bA_0\|_0$.
Suppose the coherence conditions $\gamma(\bU):=\max_i\|\mathbf{P}_U\mathbf{e_i}\| \leq \sqrt{\rho r / L}$,
$\gamma(\bV) \leq \sqrt{\rho r / T}$, and
$\xi(\bU,\bV):=\|\bU\bV'\|_\infty \leq \sqrt{\rho r / LT} $ hold for some positive constant $\rho$.
If $\mu(\Phi,\Omega)$ is sufficiently small such that the following conditions
\begin{description}
\item[$\mathcal{I}$)] $0 < \mu(\Phi,\Omega) < 1-\sqrt{\omega_{\text{max}}}$; and
\item[$\mathcal{II}$)] $(1+\alpha_{\text{max}}) \sqrt{r} \left\{ \left(\frac{1+\beta_{\text{max}}}
{1-\beta_{\text{max}}} \right)
\sqrt{\frac{\rho s}{L T}} + \mu(\Phi,\Omega)  \right\} < 1$
\end{description}
hold, where
\begin{align}
\omega_{\text{max}}&:= \rho r k \left( \frac{1}{L} + \frac{1}{T} \right) \nonumber\\
\alpha_{\text{max}}&:= \left[\frac{1}{(1-\mu(\Phi,\Omega))^2}-1\right]^{1/2} \nonumber\\
\beta_{\text{max}}&:= \frac{1}{(1-\mu(\Phi,\Omega))^2 ({\omega_{\text{max}}}^{-1}) - 1 } \nonumber
\end{align}
then there exists $\lambda>0$ for which the convex program (P1) with $\bR=\bI_L$
exactly recovers $\{\bX_{0},\bA_0\}$.
\end{corollary}

\noindent In Section \ref{sec:sat}, random matrices $\{\bX_0,\bA_0,\bR\}$ drawn from natural
ensembles are shown to satisfy I) and II) with high probability. In this case,
it is possible to arrive at simpler conditions (depending only
on $r$, $s$, and the matrix dimensions) for exact
recovery in the context of PCP; see Remark \ref{rem:pcp}.
Corollary \ref{corr:corr_1}, on the other hand, offers general conditions
stemming from a purely deterministic approach.

In the CS setting one has $\bX_0=\mathbf{0}_{L\times T}$, which implies
$\mu(\Phi,\Omega_R)=\xi_R(\bU,\bV)=\gamma_R(\bU)=\gamma(\bV)=0$.
As a result, Theorem~\ref{th:th_1} simply boils down to
a RIC-dependent sufficient condition for the exact recovery of $\bA_0$ as stated next.

\begin{corollary}\label{corr:corr_cs}
Consider given matrices $\bY \in \mathbb{R}^{L \times T}$ and $\bR \in \mathbb{R}^{L \times F}$ obeying
$\bY=\bR\bA_0$. Assume that the number of nonzero
elements per column of $\bA_0$ does not exceed $k$. If
\begin{align}
\delta_k(\bR) + k\theta_{1,1}(\bR) < 1 \label{eq:suf_cond_cs}
\end{align}
holds, then (P1) with $\bX=\mathbf{0}_{L\times T}$ exactly recovers $\bA_0$.
\end{corollary}

To place $\eqref{eq:suf_cond_cs}$ in context, consider normalizing the
rows of $\bR$. For such a compression matrix it is known that
$\delta_k(\bR) \leq (k-1) \theta_{1,1}(\bR)$, see e.g.,~\cite{rauhut}.
Using this bound together with~\eqref{eq:suf_cond_cs}, one arrives at the stricter
condition $k < \frac{1}{2} \left(1+ \theta_{1,1}^{-1}(\bR) \right)$.
This last condition is identical to the one reported in~\cite{donoho_elad_cs},
which guarantees the success of $\ell_1$-norm minimization in recovering sparse
solutions to under-determined systems of linear equations.
The conditions have been improved in recent works; see e.g.,~\cite{rauhut} and references therein.


\section{Proof of the Main Result}
\label{sec:proof}

In what follows, conditions are first derived under which
$\{\bX_0,\bA_0\}$ is the {\it unique} optimal solution of (P1). In
essence, these conditions are expressed in terms of certain dual
certificates. Then, Section~\ref{ssec:proof_dualcert_const} deals
with the construction of a valid dual certificate.


\subsection{Unique optimality conditions}
\label{ssec:proof_unique_optimality}
Recall the \textit{nonsmooth}
optimization problem (P1), and its Lagrangian
\begin{equation}
\cL(\bX,\bA,\bM)=\|\bX\|_{\ast}+\lambda\|\bA\|_1 + \langle \bM,\bY-\bX-\bR\bA \rangle \label{eq:lag_p1}
\end{equation}
where $\bM \in \mathbb{R}^{L \times T}$ is the matrix of dual
variables (multipliers) associated with the constraint in (P1). From
the characterization of the subdifferential for nuclear- and
$\ell_1$-norm~(see e.g.,~\cite{Boyd}), the subdifferential of the
Lagrangian at $\{\bX_0,\bA_0\}$ is given by (recall that $\bX_0=\bU\bm\Sigma\bV'$)
\begin{align}
\partial_{\bX} \cL(\bX_0,\bA_0,\bM) = \left\{\bU \bV' + \bW - \bM: \quad \|\bW\| \leq 1,\quad
\cP_{\Phi}(\bW)=\mathbf{0}_{L \times T}   \right\} \label{eq:subdiff_x}\\
\partial_{\bA} \cL(\bX_0,\bA_0,\bM) = \left\{\lambda {\rm sign}(\bA_0) + \lambda\bF - \bR' \bM: \quad \|\bF\|_{\infty}
\leq 1, \quad \cP_{\Omega}(\bF)=\mathbf{0}_{F \times T}  \right\}. \label{eq:subdiff_a}
\end{align}
The optimality conditions for (P1) assert that $\{\bX_0,\bA_0\}$
is an optimal (not necessarily unique) solution if and only if
\begin{align}
 \mathbf{0}_{F \times T} \in \partial_{\bA} \cL(\bX_0,\bA_0,\bM) \textrm{  and  }  \mathbf{0}_{L \times T} \in
\partial_{\bX} \cL(\bX_0,\bA_0,\bM).\nonumber
\end{align}
This can be shown equivalent to finding the pair $\{\bW,\bF\}$ that
satisfies: i) $\|\bW\| \leq 1,~\cP_{\Phi}(\bW)=\mathbf{0}_{L \times
T}$; ii) $\|\bF\|_{\infty} \leq 1,~\cP_{\Omega}(\bF)=\mathbf{0}_{F
\times T}$; and iii) $\lambda {\rm sign}(\bA_0) + \lambda \bF = \bR'
(\bU \bV' + \bW)$. In general, i)-iii) may hold for multiple
solution pairs. However, the next lemma asserts that a slight
tightening of the optimality conditions i)-iii) leads to a {\it
unique} optimal solution for (P1). See Appendix A for a proof.

\begin{lemma}\label{lemma:lemma_2}
Assume that each column of $\bA_0$ contains at most $k$ nonzero elements,
as well as $\mu(\Omega_R,\Phi) < 1$ and $\delta_k(\bR) < 1$. If there exists a dual
certificate $\bGamma \in \mathbb{R}^{L\times T}$ satisfying
\begin{description}
\item[C1)] $\cP_{\Phi} (\bGamma) = \bU \bV'$
\item[C2)] $\cP_{\Omega}(\bR' \bGamma) = \lambda \mathrm{sgn}(\bA_0)$
\item[C3)] $\|\cP_{\Phiper}(\bGamma)\| < 1 $
\item[C4)] $\|\cP_{\Omegaper}(\bR' \bGamma)\|_{\infty} < \lambda $
\end{description}
then $\{\bX_0,\bA_0\}$ is the unique optimal solution of (P1).
\end{lemma}

The remainder of the proof deals with the construction of a dual
certificate $\bGamma$ that meets C1)-C4). To this end, tighter
conditions [I) and II) in Theorem \ref{th:th_1}] for the existence
of $\bGamma$ are derived in terms of the incoherence parameters and
the RICs. For the special case $\bR=\bI_L$, the conditions in
Lemma~\ref{lemma:lemma_2} boil down to those in~\cite[Prop. 2]{CSPW11}
for PCP. However, the dual certificate
construction techniques used in~\cite{CSPW11} do not carry over to
the setting considered here, where a compression matrix $\bR$ is
present.


\subsection{Dual certificate construction}
\label{ssec:proof_dualcert_const}

Condition C1) in Lemma~\ref{lemma:lemma_2} implies that
$\bGamma=\bU\bV'+(\bI-\bP_U)\bX(\bI-\bP_V)$, for arbitrary
$\bX\in\mathbb{R}^{L \times T}$ (cf. Remark \ref{remark:proj}). Upon
defining $\bZ:=\bR'(\bI-\bP_U)\bX(\bI-\bP_U)$ and
$\bB_{\Omega}:=\lambda \rm {sign}(\bA_0)-\cP_{\Omega}(\bR'\bU\bV')$,
C1) and C2) are equivalent to $\cP_{\Omega}(\bZ)=\bB_{\Omega}$.

To express $\cP_{\Omega}(\bZ)=\bB_{\Omega}$ in terms of the unrestricted
matrix $\bX$, first vectorize
$\bZ$ to obtain $\vec(\bZ)=\left[(\bI-\bP_V) \otimes
\bR'(\bI-\bP_U)\right]\vec(\bX)$. Define $\bA:=(\bI-\bP_V) \otimes
\bR'(\bI-\bP_U)$ and an $s\times LT$ matrix $\bA_{\Omega}$ formed
with those $s$ rows of $\bA$ associated with those elements in
$\supp(\bA_0)$. Likewise, define $\bA_{\Omegaper}$ which collects
the remaining rows from $\bA$ such that $\bA=\mathbf{\Pi}
[\bA'_{\Omega}, \bA'_{\Omegaper}]'$ for a suitable row
permutation matrix $\mathbf{\Pi}$. Finally, let $\bb_{\Omega}$ be
the vector of length $s$ containing those elements of $\bB_{\Omega}$
with indices in $\supp(\bA_0)$. With these definitions, C1) and C2) can
be expressed as $\bA_{\Omega} \vec(\bX) = \bb_{\Omega}.$

To upper-bound the left-hand side of C3) in terms of $\bX$, use the
assumption $\bR\bR'=\bI_L$ to arrive at
\begin{align}
\|\cP_{\Phiper}(\bGamma)\|&{}={}\|\bR'(\bI-\bP_U)\bX(\bI-\bP_V)\|
\leq\|\bR'(\bI-\bP_U)\bX(\bI-\bP_V)\|_F=\|\bA \vec(\bX)\|. \nonumber
\end{align}
Similarly, the left-hand side of C4) can be bounded as
\begin{align}
\|\cP_{\Omegaper}(\bR' \bGamma)\|_{\infty}&{}={}\| \cP_{\Omegaper}(\bZ) +
\cP_{\Omegaper}(\bR'\bU\bV') \|_{\infty} \nonumber\\
& \leq  \| \cP_{\Omegaper}(\bZ)\|_{\infty}+ \|\cP_{\Omegaper}(\bR'\bU\bV') \|_{\infty} \nonumber\\
&=\|\bA_{\Omegaper}\vec(\bX)\|_{\infty} + \|\cP_{\Omegaper}(\bR'\bU\bV') \|_{\infty}\nonumber.
\end{align}
In a nutshell, if one can find $\bX \in \mathbb{R}^{L\times T}$ such that
\begin{description}
\item[c1)] $\bA_{\Omega} \vec(\bX) = \bb_{\Omega}$
\item[c2)] $\|\bA \vec(\bX)\| < 1$
\item[c3)] $\|\bA_{\Omegaper}\vec(\bX)\|_{\infty} + \|\cP_{\Omegaper}(\bR'\bU\bV') \|_{\infty} < \lambda$
\end{description}
hold for some positive $\lambda$, then C1)-C4) would be satisfied as
well.

The final steps of the proof entail: i) finding an appropriate
candidate solution $\hat\bX$ such that ${\rm c}1$) holds; and ii)
deriving conditions in terms of the incoherence parameters and RICs
that guarantee $\hat\bX$ meets the required bounds in ${\rm c}2$)
and ${\rm c}3$) for a range of $\lambda$ values. The following lemma
is instrumental to accomplishing i), and its proof can be found in Appendix B.

\begin{lemma}\label{lem:lemma_3}
Assume that each column of $\bA_0$ contains at most $k$ nonzero elements,
as well as $\mu(\Omega_R,\Phi) < 1$ and $\delta_k(\bR) < 1$.
Then matrix $\bA_{\Omega}$ has full row rank, and its
minimum singular value is bounded below as
\begin{align}
\sigma_{\text{min}}(\bA_{\Omega}') \geq c^{1/2}(1-\delta_k(\bR))^{1/2}
(1-\mu(\Phi,\Omega_R)).\nonumber
\end{align}
\end{lemma}
According to Lemma~\ref{lem:lemma_3}, the least-norm (LN) solution
$\hat\bX_{\rm{LN}} := \arg\min_{\bX} \left\{\|\bX\|_F^2:
\bA_{\Omega} \vec(\bX)=\bb_{\Omega} \right\}$ exists, and is given
by
\begin{equation}
\vec(\hat\bX_{\text{LN}})=\bA_{\Omega}'\left(\bA_{\Omega}\bA_{\Omega}'\right)^{-1}
\bb_{\Omega}.\label{eq:ls}
\end{equation}
\begin{remark}[Candidate dual certificate]\label{remark:dualcert}
\normalfont From the arguments at the beginning of this section, the candidate dual certificate
is $\hat\bGamma:=\bU\bV'+(\bI-\bP_U)\hat\bX_{\text{LN}}(\bI-\bP_V)$.
\end{remark}
The LN solution is an attractive choice, since it facilitates
satisfying ${\rm c}2$) and ${\rm c}3$) which require norms of
$\vec(\bX)$ to be small. Substituting the LN solution~\eqref{eq:ls} into
the left hand side of ${\rm c}2$) yields (define
$\bQ:=\bA_{\Omegaper}\bA'_{\Omega}\left(\bA_{\Omega}\bA'_{\Omega}\right)^{-1}$
for notational brevity)
\begin{align}
\|\bA\vec(\hat\bX_{\text{LN}})\|
=\left\|\left(\begin{array}{c}\bA_{\Omega}\\\bA_{\Omegaper}
\end{array}\right)\bA_{\Omega}'\left(\bA_{\Omega}\bA_{\Omega}'\right)^{-1}\bb_{\Omega}\right\|
=\left\|\left(\begin{array}{c}\bI\\\bQ \end{array}\right)\bb_{\Omega}\right\|
 \leq \left( 1+  \|\bQ\|\right)\|\bb_{\Omega}\|. \label{eq:bound_ccc2}
\end{align}
Moreover, substituting~\eqref{eq:ls} in the left hand side of ${\rm c}3$) results in
\begin{align}
\|\bQ\bb_{\Omega}\|_{\infty} + \|\cP_{\Omegaper}(\bR'\bU\bV') \|_{\infty}
\leq \|\bQ\|_{\infty,\infty}\|\bb_{\Omega}\|_{\infty} + \|\cP_{\Omegaper}(\bR'\bU\bV') \|_{\infty}.
\label{eq:bound_cc3}
\end{align}
Next, upper-bounds are obtained for $\|\bQ\|$ and $\|\bQ\|_{\infty,\infty}$; see Appendix C
for a proof.

\begin{lemma}\label{lem:lemma_4}
Assume that each column and row of $\bA_0$ contains at most $k$ nonzero elements. If
$\mu(\Omega_R,\Phi) < 1$ and $\delta_k(\bR) < 1$ hold, then
\begin{equation*}
\|\bQ\| \leq \alpha_{\rm max} := \left[\frac{1}{c(1-\delta_k(\bR))
(1-\mu (\Omega_R,\Phi))^2}-1\right]^{1/2} \label{eq:bnd_spectnorm_psudo}.
\end{equation*}
If the tighter condition I) holds instead, then
\begin{align}
\|\bQ\|_{\infty,\infty} \leq \beta_{\rm max} := \frac{\omega_{\text{max}}}
{(1-\mu (\Omega_R,\Phi))^2(1-\delta_k(\bR))
- \omega_{\text{max}} }.\nonumber
\end{align}
%
\end{lemma}
Going back to \eqref{eq:bound_ccc2}-\eqref{eq:bound_cc3}, note that
$\|\bB_{\Omega}\|_{\infty}=\|\bb_{\Omega}\|_{\infty}$ and $\|\bB_{\Omega}\|_{F}=\|\bb_{\Omega}\|$,
which can be respectively upper-bounded as
\begin{align}
\|\bB_{\Omega}\|_{\infty}{} ={}&\|\lambda {\rm sign}(\bA_0) - \cP_{\Omega}(\bR'\bU\bV')\|_{\infty}
\leq\lambda + \|\cP_{\Omega}(\bR'\bU\bV') \|_{\infty} \label{eq:bound_infnorm_b}\\
\|\bB_{\Omega}\|_F {}={} & \|\lambda {\rm sign}(\bA_0) - \cP_{\Omega}(\bR'\bU\bV')\|_F\leq
\lambda \sqrt{s} + \|\cP_{\Omega}(\bR'\bU\bV')\|_F. \label{eq:bound_normf_Bomega}
\end{align}
Finally, $\|\cP_{\Omega}(\bR'\bU\bV')\|_F$ itself can be bounded above as
\begin{align}
\|\cP_{\Omega}(\bR'\bU\bV')\|_F^2 &= |\langle \cP_{\Omega}(\bR'\bU\bV'),\cP_{\Omega}(\bR'\bU\bV') \rangle |
\stackrel{(a)}{=}|\langle \bR'\bU\bV',\cP_{\Omega}(\bR'\bU\bV') \rangle| \nonumber\\
&=|\langle \bU\bV',\bR\cP_{\Omega}(\bR'\bU\bV') \rangle|
\stackrel{(b)}{=}|\langle \cP_{\Phi}(\bU\bV'),\cP_{\Phi}(\bR\cP_{\Omega}(\bR'\bU\bV')) \rangle| \nonumber\\
&\stackrel{(c)}{\leq} \| \cP_{\Phi}(\bU\bV')\|_F \|\cP_{\Phi}(\bR\cP_{\Omega}(\bR'\bU\bV')) \|_F \nonumber\\
& \stackrel{(d)}{\leq} \|\bU\bV'\|_F \mu(\Phi,\Omega_r)\|\bR\cP_{\Omega}(\bR'\bU\bV')\|_F \nonumber\\
& \stackrel{(e)}{\leq} \sqrt{r} \mu(\Phi,\Omega_r) c^{1/2}(1+\delta_k(\bR))^{1/2} \|\cP_{\Omega}(\bR'\bU\bV')\|_F \label{eq:bound_pomegR'UV'}
\end{align}
where (a) is due to \eqref{eq:proj_exchange}, (b) follows because $\bU\bV' \in\Phi$ (thus
$\cP_{\Phi}(\bU\bV')=\bU\bV'$) and from the property in \eqref{eq:proj_exchange}.
Moreover, (c) is a direct result of the Cauchy-Schwarz inequality, while (d)
and (e) come from~\eqref{eq:mau_mat} and~\eqref{eq:rip_ineq}, respectively,
and the assumption that number of nonzero elements per column of $\bA_0$
does not exceed $k$. All in all, $\|\cP_{\Omega}(\bR'\bU\bV')\|_F \leq \sqrt{r} \mu(\Phi,\Omega_R)
c^{1/2}(1+\delta_k(\bR))^{1/2}$ and \eqref{eq:bound_normf_Bomega} becomes
\begin{align}
\|\bB_{\Omega}\|_F \leq \lambda \sqrt{s} + \sqrt{r} \mu(\Phi,\Omega_r) c^{1/2}(1+\delta_k(\bR))^{1/2}. \label{eq:bound_normf_Bomega_2}
\end{align}
Upon substituting \eqref{eq:bound_infnorm_b},
\eqref{eq:bound_normf_Bomega_2} and the bounds in Lemma
\ref{lem:lemma_4} into \eqref{eq:bound_ccc2} and
\eqref{eq:bound_cc3}, one finds that ${\rm c}2)$ and ${\rm c}3)$ hold
if there exists $\lambda>0$ such that
\begin{subequations}
\begin{align}
(1+\alpha_{\text{max}}) \left[\lambda \sqrt{s} + \sqrt{r} \mu (\Omega_R,\Phi)
c^{1/2}(1+\delta_k(\bR))^{1/2}  \right]{} < {}&1 \label{first}\\
\beta_{\text{max}} \left(\lambda + \|\cP_{\Omega}(\bR'\bU\bV')\|_{\infty} \right) +
\|\cP_{\Omegaper}(\bR'\bU\bV')\|_{\infty}{} <{}& \lambda \label{second}
\end{align}
\end{subequations}
hold. Recognizing that $\xi_R(\bU,\bV)=
\max\{\|\cP_{\Omega}(\bR'\bU\bV')\|_{\infty},$ $
\|\cP_{\Omegaper}(\bR'\bU\bV')\|_{\infty}\}$, the left-hand side of
\eqref{second} can be further bounded. After straightforward
manipulations, one deduces that conditions \eqref{first} and
\eqref{second} are satisfied for $\lambda \in
(\lambda_{\text{min}},\lambda_{\text{max}})$, where
\begin{align}
\lambda_{\text{min}}&:= \left(\frac{1+\beta_{\text{max}}}{1-\beta_{\text{max}}}\right) \xi_R(\bU,\bV)  \nonumber \\%
\lambda_{\text{max}}&:= \frac{1}{\sqrt{s}} \left[(1+\alpha_{\text{max}})^{-1} - \sqrt{r} \mu (\Omega_R,\Phi)
c^{1/2}(1+\delta_k(\bR))^{1/2}  \right].   \nonumber 
\end{align}
Clearly, it is still necessary to ensure $\lambda_{\text{max}} >
\lambda_{\text{min}}$ so that the LN solution \eqref{eq:ls} meets
the requirements ${\rm c}1$)-${\rm c}3$) [equivalently,
$\hat\bGamma$ in Remark \ref{remark:dualcert} satisfies C1)-C4) from
Lemma \ref{lemma:lemma_2}]. Condition $\lambda_{\text{max}} >
\lambda_{\text{min}}$ is equivalent to II) in
Theorem~\ref{th:th_1}, and the proof is now complete.

\begin{remark}[Satisfiability]
\normalfont From a high-level vantage point, Theorem~\ref{th:th_1} asserts that (P1) recovers
$\{\bX_0,\bA_0\}$ when the components $\bX_0$ and $\bR\bA_0$ are sufficiently incoherent,
and the compression matrix
$\bR$ has good restricted isometry properties. It should be noted though,
that given a triplet $\{\bX_0,\bA_0,\bR\}$ in general one cannot directly check whether the
sufficient conditions I) and II) hold, since e.g.,
$\delta_k(\bR)$ is NP-hard to compute~\cite{CT05}. This motivates finding a class
of (possibly random) matrices $\{\bX_0,\bA_0,\bR\}$ satisfying
I) and II), the subject dealt with next.
\end{remark}


\section{Matrices Satisfying the Conditions for Exact Recovery}
\label{sec:sat}

This section investigates triplets $\{\bX_0,\bA_0,\bR\}$ satisfying the conditions of
Theorem~\ref{th:th_1}, henceforth termed admissible matrices. Specifically,
it will be shown that low-rank, sparse, and compression matrices
drawn from certain random ensembles satisfy the sufficient conditions
of Theorem~\ref{th:th_1} with high probability.


\subsection{Uniform sparsity model}
\label{subsec:random_sparse}

Matrix $\bA_0$ is said to be generated according to the \textit{uniform sparsity} model,
when drawn uniformly at random from the collection
of all matrices with support size $s$. There is no restriction
on the amplitude of the nonzero entries. An attractive property of
this model is that it guarantees (with high probability) that
no single row or column will monopolize most nonzero entries of $\bA_0$,
for sufficiently large $\bA_0$ and appropriate scaling of the sparsity level.
This property is formalized in the following lemma
(for simplicity in exposition it is henceforth assumed that
that $\bA_0$ is a square matrix, i.e., $F=T$).

\begin{lemma}{\normalfont\cite{CSPW11}} \label{lem:lemma_8}
If $\bA_0 \in \mathbb{R}^{F\times F}$ is generated according to the uniform sparsity model
with $\|\bA_0\|_0=s$, then the maximum number $k$ of nonzero elements per column or row
of $\bA_0$ is bounded as
\begin{align}
& k \leq \frac{s}{F} \log(F)\nonumber
\end{align}
with probability higher than $1-\mathcal{O}(F^{-\zeta})$, for $s=\mathcal{O}(\zeta F)$.
\end{lemma}

In practice, it is simpler to work with the Bernoulli model
that specifies $\textrm{supp}(\bA_0)=\{(i,j):b_{i,j}=1\}$, where
$\{b_{i,j}\}$ are
independent and identically distributed (i.i.d.) Bernoulli random
variables taking value one with probability $\pi:=s/F^2$, and zero
with probability $1-\pi$. There are three important
observations regarding the Bernoulli model. First, $|\textrm{supp}(\bA_0)|$ is a
random variable, whose expected value is $s$ and matches the uniform
sparsity model. Second, arguing as in~\cite[Lemma 2.2]{CLMW09} one can
claim that  if (P1) exactly recovers  $\{\bX_0,\bA_0\}$
from data $\bY=\bX_0+\bR\bA_0$, it will also exactly
recover $\{\bX_0,\check{\bA}_0\}$ from $\check{\bY}=\bX_0+\bR\check{\bA}_0$ when
$\supp(\check{\bA}_0) \subseteq \textrm{supp}(\bA_0)$ and
the nonzero entries coincide. Third,
following the logic of \cite[Section II.C]{Candes_robustuncertainty}
one can prove that the failure rate\footnote{The
failure rate is defined as $\textrm{Pr}(\hat{\bA}\neq\bA_0)$, where $\hat{\bA}$
is the solution of (P1).} for the uniform
sparsity model is bounded by twice the failure rate corresponding to the Bernoulli
model. As a result, any recovery guarantee established for the Bernoulli model
holds for the uniform sparsity model as well.

In addition to the bound for $k$ in Lemma \ref{lem:lemma_8}, the Bernoulli model can be used
to bound $\mu(\Phi,\Omega_R)$  in terms of the
incoherence parameters $\{\gamma_R(\bU),\gamma(\bV)\}$
and the RIC $\delta_k(\bR)$. For a proof, see Appendix D.

\begin{lemma}\label{lem:lemma_5}
Let $\Lambda:=\sqrt{c(1+\delta_1(\bR))}\left[\gamma_R^2(\bU)+\gamma^2(\bV) \right]^{1/2}$
and $n:=\max\{L,F\}$. Suppose $\bA_0\in\mathbb{R}^{F\times F}$ is
generated according to the Bernoulli model with ${\rm Pr}(b_{i,j}=1)=\pi$, and
$\bR\bR'=\bI_L$. Then, there exist positive constants $C$ and $\tau$ such that
\begin{align}
\mu(\Phi,\Omega_R) \leq \sqrt{c^{-1}(1-\delta_k(\bR))^{-1}\pi} \left[C\Lambda\sqrt{\log(LF)/\pi}  + \tau
\Lambda \log(n) +1  \right]^{1/2} \label{eq:bnd_mu}
\end{align}
holds with probability at least $1-n^{-C\pi\Lambda\tau}$ if $\delta_k(\bR)$ and
the right-hand side of \eqref{eq:bnd_mu} do not exceed one.\footnote{Even
though one has $n=F$ and $\pi=s/F^2$ in the problem studied here,
Lemma~\ref{lem:lemma_5} is stated using $n$ and $\pi$ to retain generality.}
\end{lemma}

Consider~\eqref{eq:bnd_mu} when $\Lambda$ is small enough
so that the quantity inside the square brackets is close to one.
One obtains $\mu(\Phi,\Omega_R)\leq\sqrt{c^{-1}(1-\delta_k(\bR))^{-1}\pi}$,
which reduces to the bound $\mu(\Phi,\Omega)\leq\sqrt{\pi}$
derived in \cite[Section 2.5]{CLMW09} for the special case $\bR=\bI_L$.
Hence, the price paid in terms of coherence increase due
to $\bR$ is roughly $\sqrt{c^{-1}(1-\delta_k(\bR))^{-1}}>1$.
As expected,~\eqref{eq:bnd_mu} also shows that for $\bR$ with
small RICs the incoherence between subspaces $\Phi$ and $\Omega_R$ becomes
smaller, and identifiability is more likely.

The result in Lemma~\ref{lem:lemma_5} allows one to `eliminate' $\mu(\Phi,\Omega_R)$
from the sufficient conditions in Theorem~\ref{th:th_1}, which
can thus be expressed only in terms of $\{\gamma_R(\bU),\gamma(\bV),\xi_R(\bU,\bV)\}$
and the RICs of $\bR$. In the following sections, random low-rank and
compression matrices giving rise to small incoherence parameters and RICs
are described.


\subsection{Random orthogonal model}
\label{subsec:random_low_rank}

Among other implications, matrices $\bX_0$ and $\bR$ with small $\gamma_R(\bU)$ and
$\xi_R(\bU,\bV)$
are such that the columns of $\bR$ (approximately) fall outside
the column space of $\bX_0$. From a design perspective,
this suggests that the choice of an admissible $\bX_0$
(or in general an ensemble of low-rank matrices) should take
into account the structure of $\bR$, and vice versa. However, in the interest
of simplicity one could seek conditions dealing with
$\bX_0$ and $\bR$ \emph{separately}, that still ensure
$\gamma_R(\bU)$ and $\xi_R(\bU,\bV)$ are small.
This way one can benefit from the existing theory on incoherent
low-rank matrices developed in the context of matrix completion~\cite{CR08}, and matrices
with small RICs useful for CS~\cite{rauhut,Candes_robustuncertainty}.
Admittedly, the price paid is in terms of stricter conditions
that will reduce the set of admissible matrices.

In this direction, the next lemma
bounds $\gamma_R(\bU)$ and $\xi_R(\bU,\bV)$ in terms of  $\gamma(\bU):=\max_{i}\|\bP_{U}\be_i\|$,
$\gamma(\bV)$ and $\delta_k(\bR)$.

\begin{lemma}\label{lem:lemma_6}
If $\eta(\bR):=\max_{i}\|\bR\be_i\|_1/\|\bR\be_i\|$, it then holds that
\begin{align}
\gamma_R(\bU) \leq {}& {}\eta(\bR) \gamma(\bU) \label{eq:bnd_gamu}\\
\xi_R(\bU,\bV) \leq{} &{} \sqrt{c(1+\delta_1(\bR))} \eta(\bR)
\gamma(\bU) \gamma(\bV).  \label{eq:bnd_kesi}
\end{align}
\end{lemma}
\begin{IEEEproof}
Starting from the definition
\begin{align}
\gamma_R(\bU)&=\max_i \frac{\|\bP_U\bR\be_i\|}{\|\bR\be_i\|} =
\max_i \frac{\|\bP_U \sum_{\ell} \be_{\ell}\be_{\ell}' \bR\be_i\|}{\|\bR\be_i\|} \nonumber\\
& \stackrel{(a)}{\leq} \max_i  \frac{ \sum_{\ell} \|\bP_U \be_{\ell} \| |\be_{\ell}' \bR\be_i|}{\|\bR\be_i\|}
\stackrel{(b)}{\leq} \gamma(\bU) \max_i \frac{\|\bR\be_i\|_1}{\|\bR\be_i\|} \label{eq:bound_gamma_R_U}
\end{align}
where (a) follows from the Cauchy-Schwarz inequality, and (b) from the definition of
$\gamma(\bU)$.

Likewise, applying the definition of $\xi_R(\bU,\bV)$ one obtains
\begin{align}
\xi_R(\bU,\bV) &= \max_{i,j} |\be_i' \bR'\bU\bV' \be_j| \stackrel{(c)}{\leq}  \max_{i} \|\bU' \bR \be_i'\| \max_{i}
\|\bV'\be_j\| \nonumber\\
& \leq \sqrt{c(1+\delta_1(\bR))} \gamma_R(\bU) \gamma(\bV) \stackrel{(d)}{\leq} \sqrt{c(1+\delta_1(\bR))} \eta(\bR)
\gamma(\bU) \gamma(\bV)
\end{align}
where (c) follows from the Cauchy-Schwarz inequality, and (d) is due to \eqref{eq:bound_gamma_R_U}.
\end{IEEEproof}

The bounds \eqref{eq:bnd_gamu} and \eqref{eq:bnd_kesi} are proportional to
$\gamma(\bU)$ and $\gamma(\bV)$. This prompts one to consider incoherent
rank-$r$ matrices $\bX_0=\bU\bSigma\bV'$ generated from the \textit{random orthogonal}
model, which is specified as follows.
The singular vectors forming the columns of $\bU$ and $\bV$ are drawn uniformly at random from
the collection of rank-$r$ partial isometries in $\mathbb{R}^{L \times r}$
and $\mathbb{R}^{F \times r}$, respectively. There is no need for
$\bU$ and $\bV$ to be statistically independent, and no restriction in placed on the singular values
in the diagonal of $\bm\Sigma$. The adequacy of the random orthogonal model in generating
incoherent low-rank matrices is justified by the following lemma (recall $T=F\geq L$).

\begin{lemma}{\normalfont\cite{CSPW11}} \label{lem:lemma_7}
If $\bX_0 = \bU \bSigma \bV' \in \mathbb{R}^{L \times F}$ is generated according
to the random orthogonal model with $\textrm{rank}(\bX_0)=r$, then
\begin{align}
\max\{\gamma(\bU), \gamma(\bV)\}  \leq  \sqrt{ \frac{\max\{r,\log(F)\}}{F} }   \nonumber
\end{align}
with probability exceeding $1-\mathcal{O}(F^{-3} \log(F))$.
\end{lemma}


\subsection{Random compressive matrices}
\label{subsec:compression_matrix}

With reference to Lemma~\ref{lem:lemma_6} [cf. \eqref{eq:bnd_gamu} and \eqref{eq:bnd_kesi}],
it is clear that an incoherent $\bX_0$ alone may not
suffice to yield small $\gamma_R(\bU)$ and $\xi_R(\bU,\bV)$. In addition,
$\eta(\bR)\in[1,\sqrt{L}]$ should be as close as possible to one.
This can be achieved e.g.,
when $\bR$ is sparse across each column. Note that the lower bound of unity
is attained when $\bR$ has at most a single nonzero element per column, as it
is the case when $\bR=\bI_L$.

The aforementioned observations motivate considering
block-diagonal compression matrices $\bR\in\mathbb{R}^{L\times F}$,
consisting of blocks $\{\bR_i \in \mathbb{R}^{\ell \times f}\}$ where $\ell \leq f$. The
number of blocks is $n_b:=F/f$ assuming that $f$ divides $F$. The
$i$-th block is generated according to the \emph{bounded orthonormal} model as follows; see
e.g.,~\cite{rauhut}. For some positive constant $K$, (deterministically) choose a unitary matrix
$\mathbf{\Psi} \in \mathbb{R}^{f \times f}$ with bounded entries
\begin{align}
\max_{(t,k) \in \mathcal{F}\times \mathcal{F}} |\mathbf{\Psi}_{t,k}| \leq K \label{eq:boundedness}
\end{align}
where $\mathcal{F}:=\{1,...,f\}$. For each $i=1,\ldots,n_b$ form
$\bR_i:= \mathbf{\Theta}_{T^{(i)}} \mathbf{\Psi}$,
where $\mathbf{\Theta}_{T^{(i)}}:=[\mathbf{e}_{t_1^{(i)}}, \ldots,
\mathbf{e}_{t_\ell^{(i)}}]'\in\mathbb{R}^{\ell\times f}$
is a random row subsampling matrix that selects the rows of $\bm{\Psi}$ indexed
by $\mathcal{T}^{(i)}:=\{t_1^{(i)},...,t_{\ell}^{(i)}\} \subset \mathcal{F}$.
In words, $\mathbf{\Theta}_{T^{(i)}}$ is formed by those $\ell$ rows of $\mathbf{I}_f$
indexed by $\mathcal{T}^{(i)}$. The row indices in
$\mathcal{T}^{(i)}$ are selected independently at random, with uniform
probability $1/f$ from $\mathcal{F}$.
By construction, $\bR_i\bR_i'=\bI_{\ell}, i=1,\ldots,n_b$,
which ensures $\bR\bR'=\bI_L$ as required by
Theorem~\ref{th:th_1}. Most importantly, the next lemma
states that such a construction for $\bR_i$ leads to
small RICs with high probability; see e.g.,~\cite{rauhut} for the proof.

\begin{lemma}{\normalfont\cite{rauhut}} \label{lem:lemma_9}
Let $\bR_i \in \mathbb{R}^{\ell \times f}$ be generated according to the
bounded orthonormal model.
If for some $k_i\in[1,f]$, $\epsilon \in (0,1)$ and $\mu \in(0,1/2]$ the following condition
\begin{align}
\frac{\ell}{\log(10\ell)} \geq D K^2 \mu^{-2} s \log^2(100 k_i) \log(4f) \log(7 \epsilon^{-1}) \label{eq:rip_bound}
\end{align}
holds where the constant $D \leq 243,150$, then $\delta_{k_i}(\bR_i) \leq \mu$
with probability greater than $1-\epsilon$.
\end{lemma}

\noindent Lemma~\ref{lem:lemma_9} asserts that for large enough
$\ell$, the RIC $\delta_{k_i}(\bR_i)=\mathcal{O}(\log(100k_i)\log(10\ell)
\log(4f)^{1/2} \sqrt{k_i/\ell})$ with
overwhelming probability.

Let $k_i$ denote the maximum number of nonzero elements per `trimmed' column of $\bA_0$, the trimming
being defined by the block of rows of $\bA_0$ that are multiplied
by $\bR_i$ when carrying out the product $\bR\bA_0$.
With these definitions, the RIC of $\bR$ is
bounded as $\delta_k(\bR) \leq \max_i\{\delta_{k_i}(\bR_i)\}$.
For $\delta_k(\bR)$ to be small as required by Theorem~\ref{th:th_1}, the $k_i$ should be much
smaller than $\ell$. Since $\bA_0$ is generated according to the
uniform sparsity model outlined
in Section~\ref{subsec:random_sparse}, its nonzero elements are uniformly
spread across rows and columns as per Lemma~\ref{lem:lemma_8}.
Formally, it holds that $k_i \leq \kappa := (s/F n_b)\log(F n_b)$ with probability
$1-\mathcal{O}([F n_b]^{-\zeta})$, where $s=\|\bA_0\|_0=\zeta F n_b $;
see e.g.,~\cite{Bollobas}. Accordingly, from Lemma~\ref{lem:lemma_9} one can
infer that $\delta_k(\bR) = \mathcal{O}( \log(100\kappa)\log(10\ell) \log(4f)^{1/2}
\sqrt{\kappa/\ell})$ with high probability.
Note that the bound for $\delta_k(\bR)$ depends on
$k$ through the variable $s$ in $\kappa$, and the relationship
between $s$ and $k$ in Lemma~\ref{lem:lemma_8}. Regarding the RIC
$\theta_{1,1}(\bR)$, it is bounded as $\theta_{1,1}(\bR) \leq \delta_{2}(\bR)$~\cite{CT05}.
The normalization constant $c$ in \eqref{eq:rip_ineq} and \eqref{eq:theta}
also equals $L/F\ll 1$. Recalling $\eta(\bR)$ (cf. Lemma~\ref{lem:lemma_6})
which was subject of the initial discussion in this section,
it turns out that for such a construction of $\bR$ one obtains $\eta(\bR)\leq\sqrt{\ell}\ll\sqrt{L}$.
\begin{remark}[Row and column permutations]\label{rem:row_col}
\normalfont The class of admissible compression matrices can be
extended to matrices which are block diagonal
up to row and column permutations. Let $\bm\Pi_r$ ($\bm\Pi_c$) denote,
respectively,
the row (column) permutation matrices that render $\bR$ block diagonal.
Instead of~\eqref{eq:model} consider $\mathbf{\Pi}_r \bY=\mathbf{\Pi}_r\bX_0 + \mathbf{\Pi}_r \bR
\mathbf{\Pi}_c \mathbf{\Pi}_c'\bA_0$ and note that $\mathbf{\Pi}_r\bX_0$
has the same coherence parameters as $\bX_0$, while
$\mathbf{\Pi}_r \bR \mathbf{\Pi}_c$ has the same RICs as $\bR$, and $\mathbf{\Pi}_c'\bA_0$ is
still uniformly sparse. Thus, one can feed the transformed data to (P1)
and since $\mathbf{\Pi}_r$ and $\mathbf{\Pi}_c$ are invertible, $\{\bX_0,\bA_0\}$
can be readily obtained from the recovered $\{\bm\Pi_r\bX_0,\bm\Pi_c'\bA_0\}$.
\end{remark}


\subsection{Closing the loop}
\label{subsec:closing_loop}

According to Lemmata~\ref{lem:lemma_5} and~\ref{lem:lemma_6}, the incoherence
parameters $\mu(\Phi,\Omega_R)$, $\gamma_R(\bU)$
and $\xi_R(\bU,\bV)$ which play a critcal role toward exact
decomposability in Theorem~\ref{th:th_1}, can be upper-bounded in terms of
$\gamma(\bU)$ and $\gamma(\bV)$. For random matrices $\{\bX_0,\bA_0,\bR\}$
drawn from specific ensembles, Lemmata~\ref{lem:lemma_8},~\ref{lem:lemma_7}
and~\ref{lem:lemma_9} assert that the incoherence parameters $\gamma(\bU)$ and
$\gamma(\bV)$ as well as the RICs $\delta_k(\bR)$
and $\theta_{1,1}(\bR)$, are bounded above in terms of $r=\textrm{rank}(\bX_0)$,
the degree of sparsity $s=\|\bA_0\|_0$, and the underlying matrix dimensions $L,F,\ell,f$.
Alternative sufficient conditions for exact recovery,
expressible only in terms of the aforementioned basic parameters, can be obtained
by combining the bounds of this section along with
I) and II) in Theorem~\ref{th:th_1}.
Hence, in order to guarantee that (P1) recovers $\{\bX_0,\bA_0\}$ with high probability
and for given matrix dimensions, it suffices to check feasibility of a set of
inequalities in $r$ and $s$.

To this end, focus on the asymptotic case where $L$ and $F$ are large enough,
while $F=T$ for simplicity in exposition.
Recall the conditions of Theorem~\ref{th:th_1} and
suppose $\delta_k(\bR) = o(1)$ and $\mu(\Phi,\Omega_R) = o(1)$. This results in
$\alpha_{\rm max}\approx\sqrt{F/L}$ and
$\beta_{\rm max}\approx(\omega_{\rm max}^{-1}-1)^{-1}$ when $L \ll F$. Satisfaction
of I) and II) then requires $\mathcal{O}(1)$ summands in the left-hand side of II), which gives rise to
$\xi_R(\bU,\bV) = \mathcal{O}(\sqrt{L/Fs})$, $\mu(\Phi,\Omega_R)=\mathcal{O}(\sqrt{L/Fr})$, and
$\omega_{\rm max}=\mathcal{O}(1) < 1$. The latter which is indeed the
bottleneck constraint can be satisfied if $\theta_{1,1}(\bR)=\mathcal{O}(1/k)$,
$\theta_{1,1}(\bR) \gamma^2(\bV)=\mathcal{O}(1/s)$,
$\gamma_R^2(\bU)=\mathcal{O}(1/k)$,
$\gamma^2(\bV)=\mathcal{O}(1/k)$, and $\gamma_R^2(\bU)\gamma_R^2(\bV)=\mathcal{O}(1/s)$.
Utilizing the bounds in Lemmata~\ref{lem:lemma_5}--\ref{lem:lemma_9} establishes
the next corollary.

\begin{corollary} \label{corr:corr_2}
Consider given matrices $\bY\in\mathbb{R}^{L \times F}$ and
$\bR\in\mathbb{R}^{L \times F}$ obeying
$\bY=\bX_{0}+\bR\bA_0$, where
$r:=\text{rank}(\bX_{0})$ and $s:=\|\bA_0\|_0$.
Suppose that: (i) $\bX_0$ is generated according to the random orthogonal
model; (ii) $\bA_0$ is generated according to the uniform sparsity model; and (ii)
$\bR=\textrm{bdiag}(\bR_1,\ldots,\bR_{n_b})$ with blocks $\bR_i\in\mathbb{R}^{\ell\times f}$
generated according to the bounded orthogonal model.
Define $\tilde{r}:=\max\{r,\log(F)\}$. If $r$ and $s$ satisfy
\begin{description}
\item[i)] $\tilde{r} \precsim \frac{F}{\ell}$
\item[ii)] $s \precsim \min\left\{ \frac{F^2} {\ell \log(F) \tilde{r}}, \frac{F^2}{\tilde{r}^2},
\frac{F\sqrt{\ell}}{\log(10\ell) \log^{1/2}(4f) \tilde{r}}\right\}$
\item[iii)] $s^{1/2} \log\left( 100 \frac{sf}{F^2} \log\left(\frac{F^2}{f}\right)   \right)  \prec   \left[
\frac{F^2 \ell}{f \log(F^2/f) \log^2(f)}  \right]^{1/2}$
\end{description}
there is a positive $\lambda$ for which (P1) recovers $\{\bX_0,\bA_0 \}$ with high probability.
\end{corollary}

\begin{remark}[Principal components pursuit]\label{rem:pcp}
\normalfont For PCP where $\bR=\bI_L$ and $L=T$ (cf.
Corollary~\ref{corr:corr_1}), it can be readily verified that
$s\min\{r,\log(L)\}=\mathcal {O}(L^2/\log(L))$ suffices for exact recovery of
$\{\bX_0,\bA_0\}$ by solving (P1). This guarantee is of course valid with high probability, provided
$\{\bX_0,\bA_0,\bR\}$ are drawn from the random matrix ensembles outlined
throughout this section.
However, in the presence of the compression matrix $\bR$ more stringent
conditions are imposed on the rank and
sparsity level, as stated in Corollary~\ref{corr:corr_2}.
This is mainly because of the dominant summand $[\sqrt{2}k+s\gamma^2(V)]\theta_{1,1}(\bR)$
in $\omega_{\rm max}$ (cf. Theorem~\ref{th:th_1}),
which limits the extent to which $r$ and $s$ can be increased.
If the correlation between any two columns of $\bR$ is small, then higher
rank and less sparse matrices can be exactly recovered.
\end{remark}


\section{Algorithms}
\label{sec:alg}

This section deals with iterative algorithms to solve the non-smooth
convex optimization problem (P1).


\subsection{Accelerated proximal gradient (APG) algorithm}\label{subsec:proximal}

The class of accelerated proximal gradient algorithms were originally
studied in~\cite{nesterov83,nesterov05},
and they have been popularized for $\ell_1$-norm regularized regression; mostly due to
the success of the fast iterative shrinkage-thresholding algorithm (FISTA)~\cite{fista}.
Recently, APG algorithms have been applied to matrix-valued problems such as those arising
with nuclear-norm regularized estimators for matrix completion~\cite{nuclear_proximal},
and for (stable) PCP~\cite{zlwcm10,rpca_proximal}.
APG algorithms offer several attractive features, most notably a convergence
rate guarantee of $\mathcal{O}(1/\sqrt{\epsilon})$ iterations
to return an $\epsilon-$optimal solution. In addition, APG algorithms are
first-order methods that scale nicely to high-dimensional problems arising
with large networks.

The algorithm developed here builds on the APG iterations
in~\cite{rpca_proximal}, proposed to solve the stable PCP problem.
One can relax the equality constraint in (P1) and instead solve
\begin{align}
{\rm (P2)} \quad \min_{\bbS} \quad \left\{ \nu \|\bX\|_{\ast} + \nu \lambda \|\bA\|_1  +
\frac{1}{2}\|\bbY-\bbX-\bbR\bbA\|_F^2 \right\}  \nonumber
\end{align}
with $\bbS:=\left[\bbX', \bbA' \right]'$, where the least-square term penalizes violations
of the equality constraint, and $\nu > 0$ is
a penalty coefficient. When $\nu$ approaches zero,
(P2) achieves the optimal solution of (P1)~\cite{Bers}.
The gradient of $f(\bbS):=\frac{1}{2}\|\bbY-\bbX-\bbR\bbA\|_F^2$
is Lipschitz continuous with a (minimum) Lipschitz constant
$L_f=\lambda_{\max}([\bbI_L\:\:\bbR]^\prime[\bbI_L\:\:\bbR])$, i.e.,
 $\|\nabla f(\bbS_1)-\nabla f(\bbS_2)\|\leq L_f\|\bbS_1-\bbS_2\|$,
 $\forall\:\bbS_1,\bbS_2$ in the domain of $f$.

Instead of directly optimizing the cost in (P2), APG
algorithms minimize a sequence of overestimators, obtained at
judiciously chosen points $\bbT$.
Define $g(\bbS):=\nu \|\bX\|_{\ast} + \nu \lambda \|\bA\|_1$ and form the
quadratic approximation
\begin{align}\label{eq:Q_approx}
\nonumber Q(\bbS,\bbT):={}&{}f(\bbT)+\langle\nabla f(\bbT),\bbS-\bbT\rangle + \frac{L_f}{2}\|\bbS-\bbT\|_F^2+g(\bbS)\\
={}&{}\frac{L_f}{2}\|\bbS-\bbG\|_F^2+g(\bbS)+f(\bbT)-\frac{1}{2L_f}\|\nabla f(\bbT)\|_F^2
\end{align}
where $\bbG:=\bbT-(1/L_f)\nabla f(\bbT)$. With $k=1,2,\ldots$ denoting
iterations, APG algorithms generate the sequence of iterates
\begin{equation}\label{eq:apg_iterates}
\bbS[k]:=\arg\min_\bbS Q(\bbS,\bbT[k])=\arg\min_\bbS
\left\{\frac{L_f}{2}\|\bbS-\bbG[k]\|_F^2+g(\bbS)\right\}
\end{equation}
where the second equality follows from the fact that the last two summands in \eqref{eq:Q_approx} do not depend on $\bbS$.
There are two key aspects to the success of APG algorithms. First, is the selection of
the points $\bbT[k]$ where the sequence of approximations $Q(\bbS,\bbT[k])$
are formed, since these
strongly determine the algorithm's convergence rate. The choice
$\bbT[k]=\bbS[k]+\frac{t[k-1]-1}{t[k]}\left(\bbS[k]-\bbS[k-1]\right)$, where
$t[k]=\left[1+\sqrt{4t^2[k-1]+1}\right]/2$, has been shown to significantly
accelerate the algorithm resulting in convergence rate no worse than $\mathcal{O}(1/k^2)$~\cite{fista}. The second key element
stems from the possibility of efficiently solving the sequence of subproblems
\eqref{eq:apg_iterates}.
For the particular case of (P2), note that \eqref{eq:apg_iterates} decomposes into
\begin{align}
\label{eq:X_update}\bbX[k+1]:={}&{}\arg\min_{\bbX}\left\{\frac{L_f}{2}
\|\bbX-\bbG_X[k]\|_F^2+
\nu \|\bbX\|_\ast\right\}\\
\label{eq:A_update}\bbA[k+1]:={}&{}\arg\min_{\bbA}\left\{\frac{L_f}{2}\|\bbA-
\bbG_A[k]\|_F^2+
\nu \lambda\|\bbA\|_1\right\}
\end{align}
where $\bbG[k] = [\bbG_X^\prime[k]\:\bbG_A^\prime[k]]^\prime$. Letting
$\cS_{\tau}(\bM)$ with $(i,j)$-th entry given by $\textrm{sign}(m_{i,j})
\max\{|m_{i,j}|-\tau,0\}$ denote the soft-thresholding
operator, and $\bbU\bm\Sigma\bbV^\prime=\textrm{svd}(\bbG_X[k])$ the singular
value decomposition of matrix $\bbG_X[k]$,
it follows that (see, e.g.~\cite{rpca_proximal})
\begin{equation}
\bbX[k+1]=\bbU \mathcal{S}_{\frac{ \nu}{L_f}}[\bm\Sigma]\bbV^\prime,\quad
\bbA[k+1]=\mathcal{S}_{\frac{\lambda\nu}{L_f}}[\bbG_A[k]].
\end{equation}
A continuation technique is employed to speed-up convergence
of the APG algorithm. The penalty parameter $\nu$ is
initialized with a large value $\nu_0$,
and is decreased geometrically until it reaches the target value of $\bar{\nu}$.
The APG algorithm is tabulated as Algorithm~\ref{tab:APG}. Similar
to~\cite{rpca_proximal} and~\cite{nuclear_proximal}, the iterations terminate whenever
the norm of
\begin{equation*}
\bbZ[k+1]:=
\left[\begin{array}{c}L_f(\bbT_X[k]-\bbX[k+1])+(\bbX[k+1]+\bbR\bbA[k+1]-
\bbT_X[k]-\bbR\bbT_A[k])\\
L_f(\bbT_A[k]-\bbA[k+1])+\bbR^{\prime}(\bbX[k+1]+\bbR\bbA[k+1]-\bbT_X[k]-\bbR
\bbT_A[k])\end{array}\right]
\end{equation*}
drops below some prescribed tolerance, i.e., $\|\bbZ[k+1]\|_F\leq
\textrm{tol}\times \max(1,L_f\|\bbX[k]\|_F)$.
As detailed in~\cite{nuclear_proximal}, the quantity $\|\bbZ[k+1]\|_F$ upper
bounds the distance between the origin and the set of subgradients of
the cost in (P2), evaluated at $\bbS[k+1]$.

Before concluding this section, it is worth noting that
Algorithm~\ref{tab:APG} has good convergence performance, and quantifiable
iteration complexity as asserted in the following proposition adapted
from~\cite{rpca_proximal,fista}.

\begin{proposition}{\normalfont\cite{rpca_proximal}}
Let $h(.)$ and $\{\bar\bA,\bar\bX\}$ denote, respectively, the cost and an optimal solution of (P2) when $\nu:=\bar \nu$.
For $k > k_0:=\frac{\log(\nu_0/{\bar \nu})}{\log(1/\upsilon)}$, the
iterates $\{\bA[k],\bX[k]\}$ generated by Algorithm~\ref{tab:APG}
satisfy
\begin{equation*}
|h(\bA[k],\bX[k])-h(\bar\bA,\bar\bX)| \leq
\frac{4(\|\bA[k_0]-\bar\bA\|_F^2+\|\bX[k_0]-\bar\bX\|_F^2)}
{(k-k_0+1)^2}. \label{eq:iter_complx}
\end{equation*}
\end{proposition}

\begin{algorithm}[t]
\caption{: APG solver for (P1)} \small{
\begin{algorithmic}
	\STATE \textbf{input} $\bY, \bR, \lambda, \upsilon, \nu_0, \bar{\nu}
L_f=\lambda_{\max}([\bbI_L\:\:\bbR]^\prime[\bbI_L\:\:\bbR])$
    \STATE \textbf{initialize} $\bbX[0]=\bbX[-1]=\mathbf{0}_{L\times T}$,
    $\bbA[0]=\bbA[-1]=\mathbf{0}_{F\times T}$, $t[0]=t[-1]=1$, and set $k=0$.
    \WHILE {not converged}
        \STATE
$\bbT_X[k]=\bbX[k]+\frac{t[k-1]-1}{t[k]}\left(\bbX[k]-\bbX[k-1]
        \right)$.
        \STATE
$\bbT_A[k]=\bbA[k]+\frac{t[k-1]-1}{t[k]}\left(\bbA[k]-\bbA[k-1]\right)
$.
        \STATE
$\bbG_X[k]=\bbT_X[k]+\frac{1}{L_f}\left(\bbY-\bbT_X[k]-\bbR\bbT_A[k]
\right)$.
        \STATE
$\bbG_A[k]=\bbT_A[k]+\frac{1}{L_f}\bbR^\prime\left(\bbY-\bbT_X[k]-
        \bbR\bbT_A[k]\right)$.
        \STATE $\bbU\bm\Sigma\bbV^\prime=\textrm{svd}(\bbG_X[k])$,
        $\quad\bbX[k+1]=\bbU\mathcal{S}_{\nu[k]/L_f}(\bm\Sigma)
        \bbV^\prime$.
        \STATE $\bbA[k+1]=\mathcal{S}_{\lambda \nu[k]/L_f}(\bbG_A[k]).$
        \STATE $t[k+1]=\left[1+\sqrt{4t^2[k]+1}\right]/2$
        \STATE $\nu[k+1]=\max\{\upsilon \nu[k], \bar{\nu} \}$

        \STATE $k\leftarrow k+1$
    \ENDWHILE
    \RETURN $\bX[k]$, $\bbA[k]$
\end{algorithmic}}
\label{tab:APG}
\end{algorithm}


\subsection{Alternating-direction method of multipliers (AD-MoM) algorithm}
\label{subsec:admom}

The AD-MoM is an iterative augmented Lagrangian method especially
well-suited for parallel processing~\cite{Bertsekas_Book_Distr}, which has been
proven successful to tackle the optimization tasks
encountered e.g., in statistical learning
problems~\cite{mateos_dlasso},~\cite{boyd_monograph_admom}.
While the AD-MoM could be directly applied to (P1), $\bR$ couples
the entries of $\bA$ and it turns out this yields more difficult $\ell_1$-norm
minimization subproblems per iteration. To overcome this challenge, a common
technique is to introduce an auxiliary
(decoupling) variable $\bB$, and formulate the following optimization problem
\begin{align}
\text{(P3)} \quad \min_{\{\bX,\bA,\bB\}}& \|\bX\|_{*}+\lambda\|\bA\|_1 \nonumber\\
\text{s. to}&~\bY = \bX+\bR\bB  \label{eq:cons_1_p3} \\
&~\hspace{0.5mm} \bB=\bA   \label{eq:cons_2_p3}
\end{align}
which is equivalent to (P1). To tackle (P3), associate Lagrange
multipliers $\tilde{\bM}$ and $\bar{\bM}$ with the
constraints~\eqref{eq:cons_1_p3} and~\eqref{eq:cons_2_p3}, respectively.
Next, introduce the quadratically {\it augmented} Lagrangian function
\begin{align}
\cL(\bX,\bA,\bB,\tilde{\bM},\bar{\bM})= &\|\bX\|_{*} + \lambda \|\bA\|_1 +
\langle \tilde{\bM},\bB-\bA\rangle + \langle
\bar{\bM},\bY-\bX-\bR\bB\rangle \nonumber \\ &
+ \frac{c}{2}\|\bY - \bX - \bR\bB\|_{F}^{2} + \frac{c}{2} \|\bA-\bB\|_{F}^{2} \label{eq:lag_admom}
\end{align}
where $c$ is a positive penalty coefficient. Splitting the primal variables
into two groups $\{\bX,\bA\}$ and $\{\bB\}$, the AD-MoM solver entails
an iterative procedure comprising three steps per iteration $k = 1, 2,\ldots$

\begin{description}
\item [{\bf [S1]}] \textbf{Update dual variables:}
\begin{align}
    \tilde{\bM}[k]&=\tilde{\bM}[k-1]+c(\bB[k]-\bA[k]) \label{eq:multi_M}\\
    \bar{\bM}[k]&=\bar{\bM}[k-1]+ c(\bY - \bX[k]- \bR\bB[k])
\end{align}

\item [{\bf [S2]}]  \textbf{Update first group of primal variables:}
    \begin{align}
    \bX[k+1]{}={}&
    \mbox{arg}\:\min_{\bX} \left\{ \frac{c}{2}\|\bY - \bX - \bR\bB[k]\|_{F}^{2} -
    \langle \bar{\bM}[k],\bX \rangle + \|\bX\|_{*}
\right\}.
    \label{eq:S2_ADMOM_1}\\
    \bA[k+1]{}={}&
    \mbox{arg}\:\min_{\bA} \left\{ \frac{c}{2} \|\bA-\bB[k]\|_{F}^{2} -
    \langle \tilde{\bM}[k],\bA\rangle +  \lambda \|\bA\|_1
\right\}.
    \label{eq:S2_ADMOM_2}
    \end{align}

\item [{\bf [S3]}]  \textbf{Update second group of primal variables:}
        \begin{equation}
    \bB[k+1]=
    \mbox{arg}\:\min_{\bB} \left\{ \frac{c}{2}\|\bY - \bX[k+1] - \bR\bB\|_{F}^{2} +
    \frac{c}{2} \|\bA[k+1]-\bB\|_{F}^{2} -
\langle \bR'\bar{\bM}[k]-\tilde{\bM}[k],\bB \rangle \right\}
    \label{S3_ADMOM}
        \end{equation}
\end{description}
This three-step procedure implements a block-coordinate descent
on the augmented Lagrangian, with dual variable updates. The
minimization~\eqref{eq:S2_ADMOM_1} can be recast
as~\eqref{eq:X_update}, hence $\bX[k+1]$ is iteratively updated
through singular value thresholding.
Likewise, \eqref{eq:S2_ADMOM_2} can be put in the form \eqref{eq:A_update}
and the entries of $\bA[k+1]$ are updated via parallel soft-thresholding
operations.
Finally, \eqref{S3_ADMOM} is a strictly convex unconstrained quadratic program,
whose closed-form solution is obtained as the root of the linear equation
corresponding to the first-order condition for optimality.
The AD-MoM solver is tabulated under Algorithm~\ref{tab:ADMoM}. Suitable termination
criteria are suggested in~\cite[p. 18]{boyd_monograph_admom}.

Conceivably, $F$ can be quite large, thus inverting the $F\times F$ matrix
$\bR^\prime \bR + \bI_F$ to update $\bB[k+1]$ could be complex
computationally. Fortunately, the inversion needs to be carried
out once, and can be performed and cached off-line.
In addition, to reduce the inversion cost, the SVD of the compression matrix
$\bR=\bU_{R} \bSigma_{R} \bV_{R}^\prime$ can be obtained first,
and the matrix inversion lemma can be subsequently employed
to obtain $[\bR^\prime \bR + \bI_F]^{-1}=\left[\bI_L - \bV_{R} \bC
\bV_{R}^\prime\right]$, where $\bC:=\mathrm{diag}
\left(\frac{\sigma_1^2}{1+\sigma_1^2},...,\frac{\sigma_L^2}{1+\sigma_p^2}
\right)$ and $p= {\rm rank} (\bR)\ll F$. Finally, note that the AD-MoM algorithm
converges to the global optimum of the convex program (P1) as stated in
the next proposition.

\begin{proposition}{\normalfont\cite{Bertsekas_Book_Distr}}
For any value of the penalty coefficient $c>0$, the
iterates $\{\bX[k],\bA[k]\}$ converge to the optimal solution of (P1) as $k \rightarrow \infty$.
\end{proposition}

\begin{algorithm}[t]
\caption{: AD-MoM solver for (P1)} \small{
\begin{algorithmic}
	\STATE \textbf{input} $\bY, \bR, \lambda, c$
    \STATE \textbf{initialize} $\bX[0]=\bar{\bM}[-1]=\mathbf{0}_{L\times T}$,
        $\bbA[0]=\bbB[0]=\tilde{\bM}[-1]=\mathbf{0}_{F\times T}$, and set $k=0$.
    \WHILE {not converged}

        \STATE \textbf{[S1] Update dual variables:}

        \STATE $\tilde{\bM}[k]=\tilde{\bM}[k-1]+c(\bB[k]-\bA[k])$

        \STATE $\bar{\bM}[k]=\bar{\bM}[k-1]+c(\bY - \bX[k] - \bR\bB[k])$

        \STATE \textbf{[S2] Update first group of primal variables:}
        \STATE $\bbU\bm\Sigma\bbV^\prime=\textrm{svd}(\bY-\bR \bA[k] + c^{-1}\bar{\bM}[k])$,
                $\quad\bbX[k+1]=\bbU\mathcal{S}_{1/c}(\bm\Sigma)
                \bbV^\prime$.
        \STATE $\bA[k+1]=c^{-1}\cS_{\lambda}(\tilde{\bM}[k]+c\bB[k])$.

        \STATE \textbf{[S3] Update second group of primal variables:}

        \STATE $\bB[k+1]=\bA[k+1]+(\bR' \bR + \bI_F)^{-1}\left[ \bR'( \bY - \bX[k+1]
                - \bR \bA[k+1]) - c^{-1} (\tilde{\bM}[k] - \bR' \bar{\bM}[k]) \right]$

		\STATE $k\leftarrow k+1$
    \ENDWHILE
    \RETURN $\bA[k],\bX[k]$
\end{algorithmic}}
\label{tab:ADMoM}
\end{algorithm}

\begin{remark}[Trade-off between stability and convergence rate] The APG algorithm exhibits
a convergence rate guarantee of $\mathcal{O}(1/{k^2})$~\cite{nesterov83},
while AD-MoM only attains $\mathcal{O}(1/k)$~\cite{cnvg_admom}.
For the problem considered here, APG needs an appropriate continuation
technique to achieve the predicted performance~\cite{rpca_proximal}.
Extensive numerical tests with Algorithm~\ref{tab:APG} suggest that the
convergence rate can vary considerably for different choices e.g., of the matrix $\bR$.
The AD-MoM algorithm on the other hand exhibits less variability in terms
of performance, and only requires tuning $c$. It is also better suited for
the constrained formulation (P1), since it does not need to resort to a relaxation.
\end{remark}


\section{Performance Evaluation}
\label{sec:sims}

The performance of (P1) is assessed in this section
via computer simulations.


\subsection{Exact recovery}
\label{subsec:exact_recovery}

Data matrices are generated according to $\bY=\bX_0+\bV_R'\bA_0$.
The low-rank component $\bX_0$ is generated
from the bilinear factorization model $\bX_0 = \bW\bZ'$, where
$\bW$ and $\bZ$ are $L\times r$
and $T \times r$ matrices with i.i.d. entries drawn from Gaussian
distributions $\mathcal{N}(0,1/L)$ and $\mathcal{N}(0,1/T)$,
respectively. Every entry of $\bA_0$ is randomly drawn from the
set $\{-1,0,1\}$ with ${\rm Pr} (a_{i,j}=-1)={\rm Pr}(a_{i,j}=1)=\pi/2$.
The columns of $\bV_R\in\mathbb{R}^{F\times L}$ comprise the
right singular vectors of the random matrix $\bR=\bU_R\bSigma_R\bV_R'$, with
i.i.d. Bernoulli entries with parameter $1/2$
(cf. Remark~\ref{rem:ortho_R}). The dimensions are $L=105$, $F=210$, and $T=420$. To demonstrate that (P1) is capable of recovering the exact values of $\{\bX_0,\bA_0\}$, the
optimization problem is solved for a wide range of values of $r$ and $s$ using the
APG algorithm (cf. Algorithm~\ref{tab:APG}).

\begin{figure}[t]
  \centering
  \centerline{\epsfig{figure=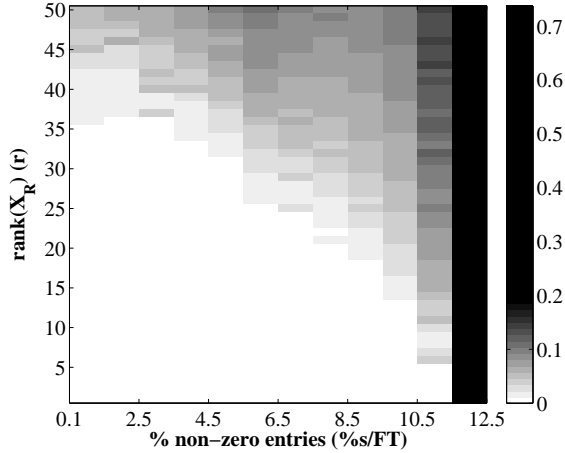,width=0.5\textwidth}}
  \caption{Relative error $e_r:=\|\bA_0-\hat{\bA}\|_F/\|\bA_0\|_F$ for various
  values of $r$ and $s$ where $L=105,~F=210$, and $T=420$. White represents
  exact recovery ($e_r\thickapprox0$), while black represents $e_r
  \thickapprox 1$. }
  \label{fig:fig_exactrecovery}
\end{figure}

\begin{table}[t]
\caption{Recovery performance by varying the size of $\bR$ when $r=10$ and $\pi=0.05$.}
\label{tab:table_2}
\begin{center}
\begin{tabular} {|c|c|c|c|c|c|}
\hline
$L$ & ${\rm rank}(\bX_0)$ & $\|\bA_0\|_0$ & ${\rm rank}(\hat{\bX})$ &
$\|\hat{\bA}\|_0$ & $\|\hat{\bA}-\bA_0\|_F/\|\bA_0\|_F$ \\
\hline
$F$ & $10$ & $4410$ & $10$ & $4419$ & $2.0809 \times 10^{-6}$  \\
\hline
$F/2$ & $10$ & $4410$ & $10$ & $4407$ & $6.4085 \times 10^{-5}$  \\
\hline
$F/3$ & $10$ & $4410$ & $10$ & $9365$ & $7.76 \times 10^{-2}$ \\
\hline
$F/5$ & $10$ & $4410$ & $14$ & $14690$ & $6.331 \times 10^{-1}$  \\
\hline
\end{tabular}
\end{center}
\end{table}

\begin{table}[h]
\caption{Performance comparison of LS-PCP and Algorithm~\ref{tab:APG}
averaged over ten random realizations}.
\label{tab:table_3}
\begin{center}
\begin{tabular} {|c|c|c|c|c|c|}
\hline
Algorithm & $r=5,~\pi=0.01$ & $r=5,~\pi=0.05$ & $r=10,~\pi=0.01$ & $r=10,~\pi=0.05$ \\
\hline
LS-PCP & $0.6901$ & $0.6975$ & $0.7001$ & $0.7023$ \\
\hline
Algorithm~\ref{tab:APG} & $7.81 \times 10^{-6}$ & $3.037 \times 10^{-5}$ &
$1.69 \times 10^{-5}$ & $6.4 \times 10^{-5}$\\
\hline
\end{tabular}
\end{center}
\end{table}

Let $\hat{\bA}$ denote the solution of (P1) for a suitable value
of $\lambda$. Fig.~\ref{fig:fig_exactrecovery} depicts the relative error in recovering
$\bA_0$, namely $\|\hat{\bA}-\bA_0\|_F/\|\bA_0\|_F$ for various values of $r$
and $s$. It is apparent that (P1) succeeds in recovering $\bA_0$ for
sufficiently sparse $\bA_0$ and low-rank $\bX_0$ from the observed data
$\bY$. Interestingly, in cases such as $s=0.1 \times FT$ or $r = 0.3 \times
{\rm min}(L,T)$ there is hope for recovery. In this example, one can exactly recover
$\{\bX_0,\bA_0\}$ when $s=0.0127 \times FT$ and $r=0.2381\times \min(L,T)$.
A similar trend is observed for the recovery of $\bX_0$, and the corresponding
plot is omitted to avoid unnecessary repetition.
For different sizes of the matrix $\bR$, performance results averaged
over ten realizations of the experiment are listed in Table~\ref{tab:table_2}.
The smaller the compression ratio $L/F$ becomes, less observations are available
and performance degrades accordingly.
In particular, the error performance degrades significantly for a challenging
instance where $L/F=0.2$ and $r=0.4 \times {\rm min}(L,F)$ (cf. the last row
of Table~\ref{tab:table_2}).

The results of~\cite{CLMW09} and~\cite{CSPW11} assert that exact recovery of
$\{\bX_0,\bA_0\}$ from the observations $\bY=\bX_0+\bA_0$ is possible under
some technical conditions. Even though the algorithms therein are not directly applicable
here due to the presence of $\bR$, one may still consider applying PCP after suitable
pre-processing of $\bY$. One possible approach is to find the LS
estimate of the superposition $\bX_0+\bA_0$ as $\hat{\bY}=\bR^{\dag}
\bY$, and then feed a PCP algorithm with $\hat{\bY}$ to obtain
$\{\bX_0,\bA_0\}$. Comparisons between (P1) and the
aforesaid two-step procedure are summarized in Table~\ref{tab:table_3}.
It is apparent that the heuristic performs very
poorly, which is mainly due to the null space of matrix $\bR$
(when $F=2L$) that renders LS estimation inaccurate.


\subsection{Unveiling network anomalies via sparsity and low rank}
\label{subsec:unveiling_anomalies}

In the backbone of large-scale networks, origin-to-destination (OD) traffic
flows experience abrupt changes which can result in congestion, and
limit the quality of service provisioning of the end users. These so-termed
\emph{traffic volume anomalies} could be due to external sources such as
network failures, denial of service attacks, or, intruders which hijack the
network services~\cite{MC03},~\cite{LCD04},~\cite{zggr05}. Unveiling such
anomalies is a crucial task towards engineering network traffic. This is a
challenging task however, since the available data are usually high-dimensional
noisy link-load measurements, which comprise the superposition of
\emph{unobservable} OD flows as explained next.

Consider a backbone  network with topology represented by the
directed graph $G(\cal{N},\cal{L})$, where $\mathcal{L}$ and $\cal{N}$ denote the set of
links and nodes (routers) of cardinality $|\mathcal{L}|=L$ and $|\calN|=N$, respectively.
The network transports $F$ end-to-end flows associated
with specific OD pairs. For backbone
networks, the number of network layer
flows is typically much larger than the number of physical links $(F \gg L)$. Single-path
routing is considered here to send the traffic flow from a source to its intended
destination. Accordingly, for a particular
flow multiple links connecting the corresponding OD pair are
chosen to carry the traffic. Sparing details that can
be found in~\cite{MMG11}, the traffic $\bbY:=[y_{l,t}]\in\mathbb{R}^{L\times T}$
carried over links $l\in\cL$ and measured at time instants $t\in [1,T],$
can be compactly expressed as
\begin{equation}
\bY=\bR \left(\bZ + \bA\right) + \bE \label{eq:Y}
\end{equation}
where the fat routing matrix $\bbR:=[r_{\ell,f}]\in\{0,1\}^{L\times F}$
is fixed and given, $\bbZ:=[z_{f,t}]$ denotes the unknown `clean'
traffic flows over the time horizon of
interest, $\bbA:=[a_{f,t}]$ collects the traffic volume anomalies
across flows and time, and $\bE:=[e_{l,t}]$ captures measurement errors.

Common temporal patterns among the traffic flows in addition
to their periodic behavior, render most rows (respectively columns)
of $\bZ$ linearly dependent, and thus $\bZ$
typically has  low rank~\cite{LCD04, rzwq11}. Anomalies are expected to occur sporadically
over time, and only last for short periods relative to the (possibly long)
measurement interval $[1,T]$. In addition, only a small fraction of the flows are
supposed to be anomalous at any given
time instant. This renders the anomaly matrix $\bA$ sparse across
rows and columns. Given link measurements $\bY$ and the routing matrix $\bR$,
the goal is to estimate $\bA$ by capitalizing on the sparsity of
$\bA$ and the low-rank property of $\bZ$. Since the primary goal is to recover
$\bA$, define $\bX:=\bR\bZ$ which inherits
the low-rank property from $\bZ$, and consider
\begin{align}
\bY= \bX + \bR \bA + \bE \label{eq:Y_modf}
\end{align}
which is identical to \eqref{eq:model} modulo small measurement
errors in $\bE\in\mathbb{R}^{L\times T}$. If $\bE=\mathbf{0}_{L\times T}$, then (P1)
can be used to unveil network anomalies, whereas (P2) is more suitable
for a noisy setting.
\begin{remark}[Distributed algorithms] Implementing Algorithms \ref{tab:APG}
and \ref{tab:ADMoM} presumes that network nodes
communicate their local link traffic measurements to a central processing
unit, which uses their aggregation in $\bY$ to determine network anomalies.
Collecting all this information can be
challenging due to excessive protocol overhead, or,
may be even impossible in e.g., wireless sensor
networks operating under stringent power budget constraints. Performing the
optimization in a centralized fashion raises robustness concerns as well, since the central
node carrying out the specific task at hand represents an isolated point of
failure. These reasons motivate devising \emph{fully-distributed} algorithms for unveiling
anomalies in large scale networks, whereby each node carries out simple
computational tasks locally, relying only on its local measurements and
messages exchanged with its directly connected neighbors. This is the subject
dealt with in an algorithmic companion paper~\cite{tsp_rankminimization_2012}, which puts
forth a general framework for in-network sparsity-regularized
rank minimization.
\end{remark}

\begin{figure}[t]
\centering
  \centerline{\epsfig{figure=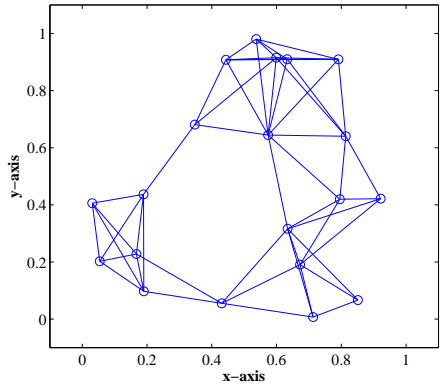,width=0.4\textwidth}}
\vspace{-5mm}\caption{Network topology graph.}
  \label{fig:fig_nettolpology}
\end{figure}

\begin{figure}[t]
\centering
\begin{tabular}{cc}
     \epsfig{file=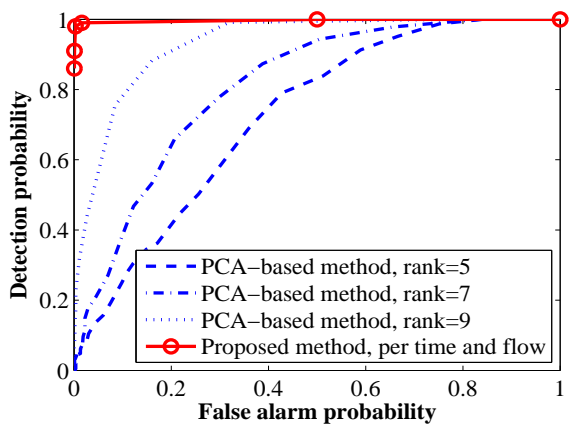,width=0.5\linewidth, height=2.3 in } &
     \epsfig{file=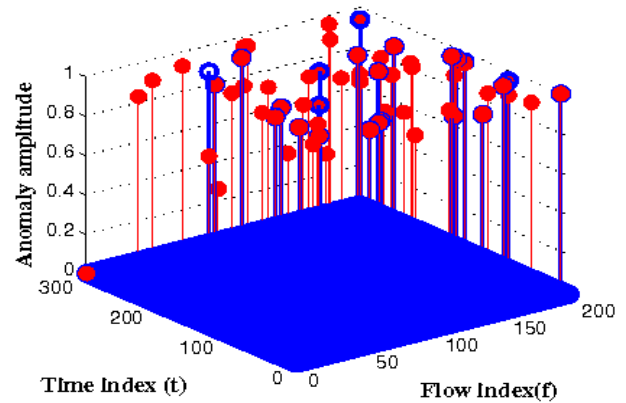,width=0.5\linewidth, height=2.3 in } \\
     (a) &
     (b) \\
  \end{tabular}
  \caption{Performance for synthetic data. (a) ROC curves of the proposed
  versus the PCA-based method with $\pi=0.001$, $r=10$ and $\sigma=0.1$. (b) Amplitude of
  the true and estimated anomalies for $P_F=10^{-4}$ and $P_D=0.97$.
Lines with open and filled circle markers denote the true and estimated anomalies, respectively.}
  \label{fig:fig_perfsynthetic}
\end{figure}

\noindent\textbf{Synthetic network data}. A network of $N=20$ agents is
considered as a realization of the random geometric graph model, that is,
agents are randomly placed on the unit square and two agents communicate with
each other if their Euclidean distance is less than a prescribed communication
range of $0.35$; see Fig.~\ref{fig:fig_nettolpology}. The network graph is bidirectional
and comprises $L=106$ links, and $F=N(N-1)=380$ OD flows. For each candidate
OD pair, minimum hop count routing is considered to form the routing matrix
$\bR$. With $r=10$, matrices $\{\bX_0,\bA_0\}$ are generated as explained in
Section \ref{subsec:exact_recovery}. With reference to~\eqref{eq:Y}, the
entries of $\bE$ are i.i.d., zero-mean, Gaussian with variance $\sigma^2$, i.e.,
$e_{l,t}\sim\mathcal{N}(0,\sigma^2)$.

\noindent\textbf{Real network data}. Real data including
OD flow traffic levels are collected from the operation of the Internet2 network
(Internet backbone network across USA)~\cite{Internet2}. OD flow
traffic levels are recorded for a three-week operation of Internet2
during Dec. 8--28, 2008~\cite{LCD04}. Internet2
comprises $N=11$ nodes, $L=41$ links, and $F=121$ flows. Given the
OD flow traffic measurements, the link loads in $\bY$ are obtained
through multiplication with the Internet2 routing matrix~\cite{Internet2}. Even though $\bY$
is `constructed' here from flow measurements, link loads can be typically
acquired from simple network management protocol (SNMP) traces~\cite{MC03}.
The available OD flows are a superposition of `clean' and anomalous traffic, i.e.,
the sum of unknown `ground-truth' low-rank and sparse matrices $\bX_{0}+\bA_0$
adhering to \eqref{eq:Y} when $\bR=\bI_L$.
Therefore, PCP is applied first to obtain an
estimate of the `ground-truth' $\{\bX_0,\bA_0\}$. The estimated $\bX_0$
exhibits three dominant singular values, confirming the low-rank property of $\bX_0$.

\noindent\textbf{Comparison with the PCA-based method.} To highlight the merits of the
proposed anomaly detection algorithm, its performance is compared with the
workhorse PCA-based
approach of~\cite{LCD04}. The crux of this method is that
the anomaly-free data is expected to be low-rank, whereas the presence of anomalies
considerably increases the rank of $\bY$. PCA requires a priori knowledge of the
rank of the anomaly-free traffic matrix, and is unable to identify anomalous flows,
i.e., the scope of~\cite{LCD04} is limited to a single anomalous flow per time slot.
Different from~\cite{LCD04}, the developed framework here enables identifying multiple
anomalous flows per time instant. To assess performance, the detection rate will be
used as figure of merit, which measures the algorithm's success in identifying anomalies
across both flows and time.

\begin{figure}[t]
\centering
\begin{tabular}{cc}
     \epsfig{file=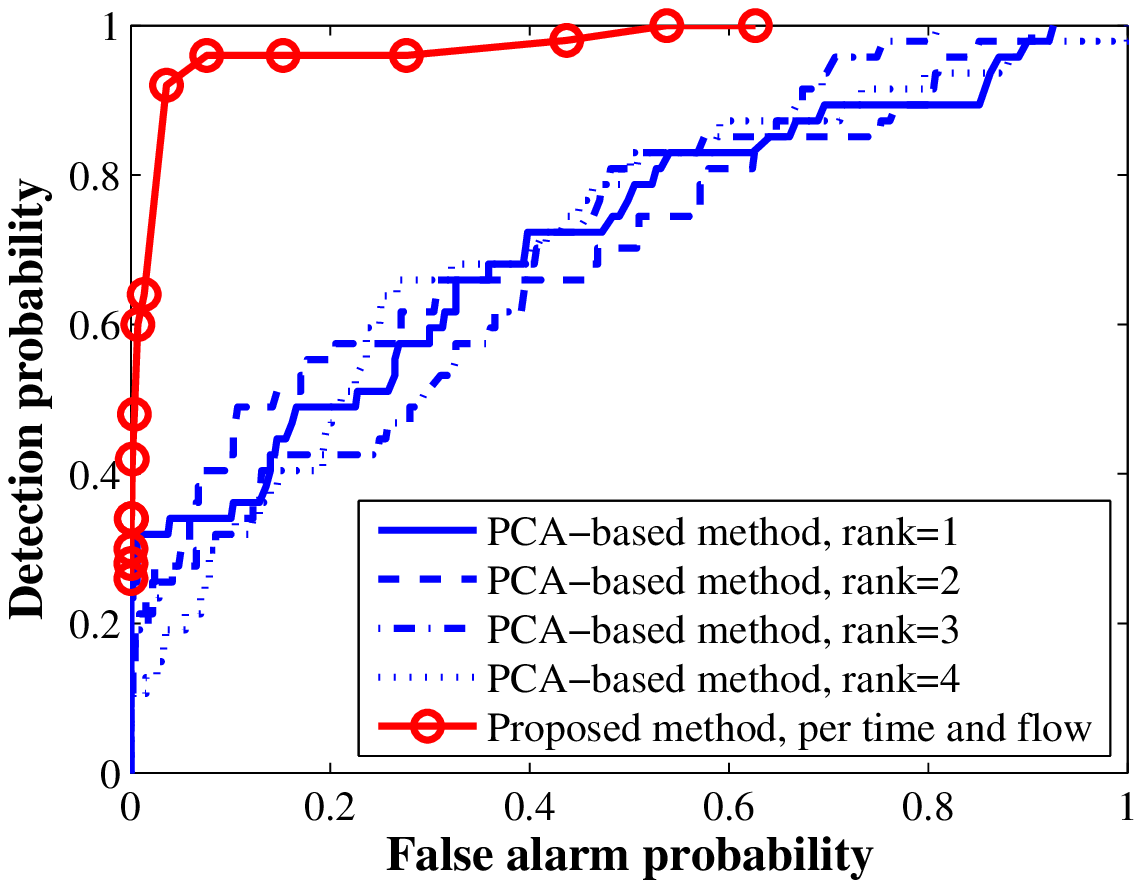,width=0.5\linewidth,height=2.3 in} &
     \epsfig{file=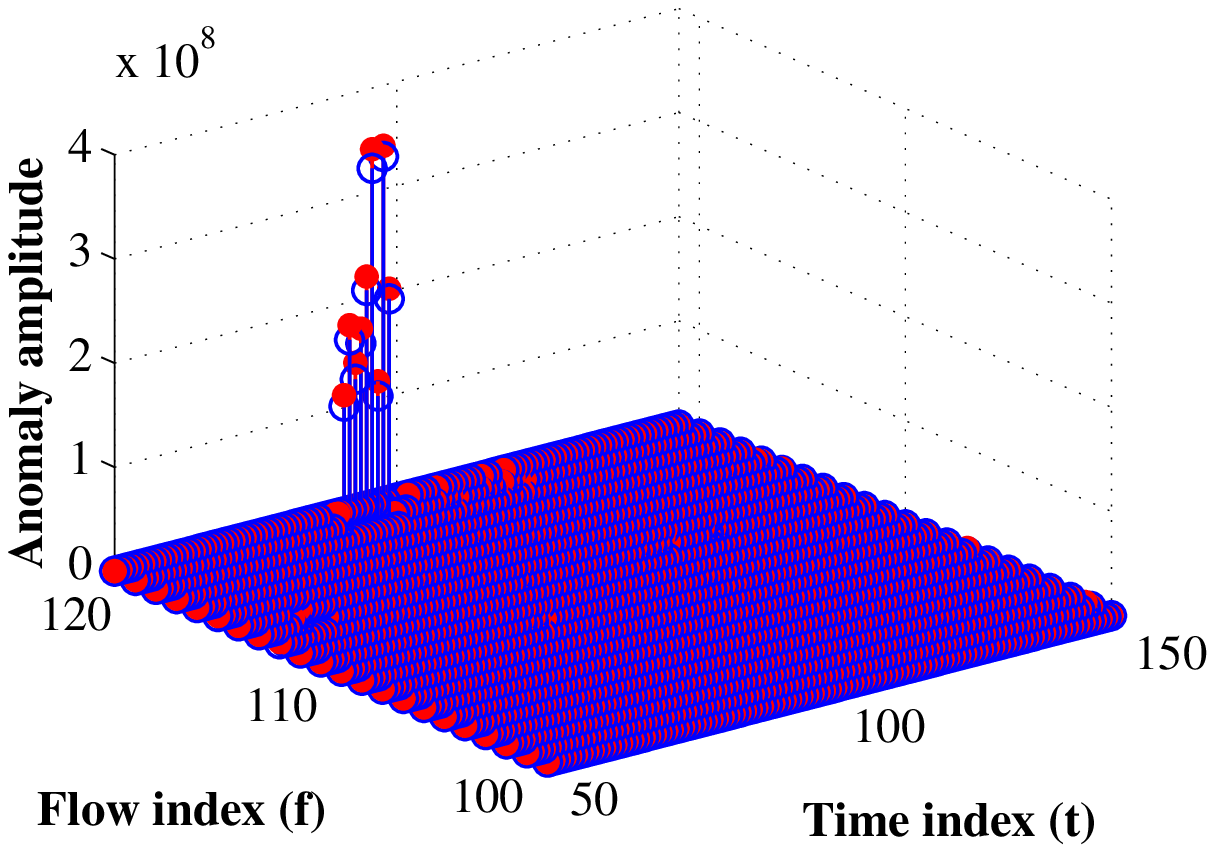,width=0.5\linewidth,height=2.3 in} \\
     (a) &
     (b) \\
  \end{tabular}
  \caption{Performance for Internet2 network data. (a) ROC curves of the proposed
versus the PCA-based method. (b)
Amplitude of the true and estimated anomalies for $P_F=0.04$ and $P_D=0.93$. Lines with open and filled circle markers
denote the true and estimated anomalies, respectively.}
  \label{fig:fig_perfreal}
\end{figure}

For the synthetic data case, ROC curves are depicted in
Fig. \ref{fig:fig_perfsynthetic} (a), for different values of
the rank required to run the PCA-based method. It is apparent that
the proposed scheme detects accurately the anomalies, even at low
false alarm rates. For the particular case of $P_F=10^{-4}$ and $P_D=0.97$, Fig.
\ref{fig:fig_perfsynthetic} (b) illustrates the magnitude of the true and
estimated anomalies across flows and time. Similar results are depicted for
the Internet2 data in Fig. \ref{fig:fig_perfreal}, where it is also apparent that the
proposed method markedly outperforms PCA in terms of detection performance.
For an instance of $P_F=0.04$ and $P_D=0.93$, Fig. \ref{fig:fig_perfreal} (b)
shows the effectiveness of the proposed algorithm in terms of unveiling
the anomalous flows and time instants.


\section{Closing Comments}
\label{sec:disscusion}

This paper deals with recovery of low-rank plus \emph{compressed}
sparse matrices via convex optimization.
The corresponding task arises with network traffic monitoring,
brain activity detection from undersampled fMRI, and video surveillance tasks, while it encompasses
compressive sampling and principal components pursuit. To estimate
the unknowns, a convex optimization program is formulated that mininimizes a trade-off between
the nuclear and $\ell_1$-norm of the low-rank and sparse components, respectively,
subject to a data modeling constraint. A deterministic approach is adopted to characterize
local identifiability and sufficient conditions for exact recovery via the aforementioned
convex program. Intuitively, the obtained conditions require: i) incoherent,
sufficiently low-rank and sparse components;
and ii) a compression matrix that behaves like an
isometry when operating on sparse vectors.
Because these conditions are in general NP-hard to check, it is shown that matrices
drawn from certain random ensembles can be recovered
with high probability.
First-order iterative algorithms are developed to
solve the nonsmooth optimization problem, which
converge to the globally optimal solution with quantifiable complexity.
Numerical tests with synthetic and real network data corroborate
the effectiveness of the novel approach in unveiling traffic anomalies across
flows and time.

One can envision several extensions to this work, which provide new and challenging directions
for future research. For instance, it seems that the requirement of an
orthonormal compression matrix is only a restriction imposed by the method
of proof utilized here. There should be room for tightening the bounds used
in the process of constructing the dual certificate, and hence obtain
milder conditions for exact recovery.
It would also be interesting to study stability of the proposed estimator
in the presence of noise and missing data.
In addition, one is naturally tempted to search for a broader class
of matrices satisfying the exact recovery conditions, including e.g., non block-diagonal
and binary routing (compression) matrices arising with the network
anomaly detection task.


{\Large\appendix}


\noindent\normalsize \emph{\textbf{A. Proof of Lemma~\ref{lemma:lemma_2}}}:
Suppose $\{\bX_0,\bA_0\}$ is an optimal
solution of (P1). For the nuclear norm and the $\ell_1$-norm at point
$\{\bX_0,\bA_0\}$ pick the subgradients $\bU\bV' + \bW_0$ and ${\rm sign}(\bA_0) + \bF_0$,
respectively, satisfying the optimality condition
\begin{equation}
\lambda {\rm sign}(\bA_0) + \lambda \bF = \bR'(\bU \bV' + \bW).\label{eq:opt_cond}
\end{equation}
Consider a feasible solution
$\{\bX_0 + \bR\bH,\bA_0-\bH\}$ for arbitrary nonzero $\bH$. The
subgradient inequality yields
\begin{align}
\|\bX_0+\bR\bH\|_{\ast} + \lambda  \|\bA_0-\bH\| \geq & \|\bX_0\|_{\ast} + \lambda \|\bA_0\|_1
+ \underbrace{ \langle \bU\bV' + \bW_0, \bR\bH \rangle - \lambda
\langle \sgn(\bA_0) + \bF_0, \bH \rangle }_{:= \varphi(\bH)}. \nonumber
\end{align}
To guarantee uniqueness, $\varphi(\bH)$ must be positive. Rearranging terms one obtains
\begin{align}
\varphi(\bH)= \langle \bW_0,\bR\bH \rangle - \lambda \langle \bF_0, \bH \rangle + \langle \bR'\bU\bV' - \lambda {\rm
sign}(\bA_0), \bH \rangle. \label{eq:gamma_h}
\end{align}
The value of $\bW_0$ can be chosen such that $\langle \bW_0, \bR\bH \rangle = \|\cP_{\Phiper}(\bR\bH)\|_{\ast}$. This
is because, $\|\cP_{\Phiper}(\bR\bH)\|_{\ast} = \sup_{\|\bar{\bW}\| \leq 1} |\langle \bar{\bW},\cP_{\Phiper}(\bR\bH)
\rangle |$, thus there exists a $\bar{\bW}$ such that
$\langle \cP_{\Phiper}(\bar{\bW}), \bR\bH \rangle =
\|\cP_{\Phiper}(\bR\bH)\|_{\ast}$. One can then choose $\bW_0:=\cP_{\Phiper}(\bar{\bW})$ since
$\|\cP_{\Phiper}(\bar{\bW})\| \leq \|\bar{\bW} \| \leq 1$ and $\cP_{\Phi}(\bW_0)=\mathbf{0}_{L \times T}$. Similarly,
if one selects $\bF_0:=-\cP_{\Omegaper}({\rm sign}(\bH))$, which satisfies $\cP_{\Omega}(\bF_0) = \mathbf{0}_{F \times T}$ and $\|\bF_0\|_{\infty} = 1$, then $\langle \bF_0, \bH \rangle = -\|\cP_{\Omegaper}(\bH)\|_1$. Now, using \eqref{eq:opt_cond}, equation~\eqref{eq:gamma_h} is expressed as
\begin{equation}
\varphi(\bH) = \|\cP_{\Phiper}(\bR\bH)\| + \lambda \|\cP_{\Omegaper}(\bH)\| + \langle \lambda \bF - \bR' \bW, \bH
\rangle. \nonumber
\end{equation}
From the triangle inequality $|\langle \lambda \bF - \bR' \bW, \bH \rangle | \leq \lambda |\langle \bF, \bH \rangle| + |\langle \bR'\bW, \bH \rangle|$, it thus follows that
\begin{align}
\varphi(\bH) \geq \left(\|\cP_{\Phiper}(\bR\bH)||_{\ast}-|\langle \bR' \bW, \bH \rangle |\right) + \lambda \left(
\|\cP_{\Omegaper}(\bH)\|_1 - |\langle \bF, \bH \rangle |\right).
\end{align}
Since $\cP_{\Phiper}(\bW) = \bW$, it is deduced that $|\langle \bW, \bR\bH \rangle | = |\langle \bW,
\cP_{\Phiper}(\bR\bH) \rangle |\leq\|\bW\| \|\cP_{\Phiper}(\bR\bH)\|_{\ast}$. Likewise, $\cP_{\Omegaper}(\bF)=\bF$
yields $| \langle \bF,\bH \rangle | = | \langle \bF, \cP_{\Omegaper}(\bH) \rangle | \leq \|\bF\|_{\infty}
\|\cP_{\Omegaper}(\bH)\|_1$. As a result
\begin{align}
\varphi(\bH) \geq& (1 - \|\bW\|) \|\cP_{\Phi}(\bR\bH)\|_{\ast} + \lambda (1 -
\|\bF\|_{\infty})\|\cP_{\Omegaper}(\bH)\|_1\nonumber\\
&\geq (1-\max\{\|\bW\|, \|\bF\|_{\infty}\})\{\|\cP_{\Phiper}(\bR\bH)\|_{\ast} +
\lambda \|\cP_{\Omegaper}(\bH)\|_1\}.
\end{align}
Now, if $\|\bW\|<1$ and $\|\bF\|_{\infty}<1$, since $\Phi \cap \Omega_R = \{\mathbf{0}_{L \times T}\}$ and $\bR\bH
\neq \mathbf{0}_{L\times T},~\forall \bH\in\Omega \backslash \{\mathbf{0}_{F\times T}\}$, there is no $\bH\in\Omega$
for which $\bR\bH \in \Phi$, and therefore, $\varphi(\bH)>0$.

Since $\bW$ and $\bF$ are related
through~\eqref{eq:opt_cond}, upon defining $\bGamma:=\bR'(\bU\bV'+\bW)$, which is indeed the dual variable for (P1),
one can arrive at conditions C1)-C4).\hfill$\blacksquare$


\noindent\normalsize \emph{\textbf{B. Proof of Lemma~\ref{lem:lemma_3}}}:
To establish that the rows of $\bA_{\Omega}$ are linearly independent, it suffices
to show that $\|\bA'\vec(\bH)\|>0$, for all nonzero $\bH\in\Omega$.
It is then possible to
\begin{align}
\|\bA'\vec(\bH)\|&=\|(\bI-\bP_V) \otimes (\bI-\bP_U)\bR \vec(\bH)\|
=\|(\bI-\bP_U)\bR\bH(\bI-\bP_V)\|_F \nonumber \\
&  =\|\cP_{\Phiper}(\bR\bH)\|_F =\|\bR\bH-\cP_{\Phi}(\bR\bH)\|_F  \nonumber\\
&  \stackrel{(a)}{\geq} \|\bR\bH\|_F-\|\cP_{\Phi}(\bR\bH)\|_F
\stackrel{(b)}{\geq} \|\bR\bH\|_F(1-\mu(\Omega_R,\Phi)) \label{eq:RC}
\end{align}
where (a) follows from the triangle inequality, and
(b) from \eqref{eq:mau_mat}. The assumption $\delta_k(\bR)<1$
along with the fact that no column of $\bH$ has more than $k$ nonzero elements, imply that $\bR\bH \neq \mathbf{0}_{L \times
T}$. Since $\mu(\Omega_r,\Phi)<1$ by assumption, the claim follows from \eqref{eq:RC}.

To arrive at the desired bound on $\sigma_{\text{min}}(\bA_{\Omega}')$,
recall the definition of the minimum singular value~\cite{Horn}
\begin{align}
\sigma_{\text{min}}(\bA_{\Omega}') &= \min_{\bH \in \Omega \backslash\{\mathbf{0}_{F \times T}\}} \frac{\|\bA'
\vec(\bH)\|}{\|\vec(\bH)\|} = \min_{\bH\in\Omega\backslash\{\mathbf{0}_{F \times T}\}}\frac{\|(\bI-\bP_{U})\bR \bH
(\bI-\bP_{V})\|_F}{\|\bH\|_F}\nonumber\\
&\stackrel{(c)}{=}\min_{\bH \in \Omega\backslash\{\mathbf{0}_{F \times T}\}} \frac{\|\bR\bH\|_F}{\|\bH\|_F} \times
\frac{\|\cP_{\Phiper}(\bR\bH)\|_F}{\|\bR\bH\|_F}
\stackrel{(d)}{\geq} c^{1/2}(1-\delta_k(\bR))^{1/2}\min_{\bZ\in\Omega_R\backslash\{\mathbf{0}_{L \times T}\}}
\frac{\|\cP_{\Phiper}(\bZ)\|_F}{\|\bZ\|_F} \nonumber\\
&=c^{1/2}(1-\delta_k(\bR))^{1/2}\min_{\bZ\in\Omega_r\backslash\{\mathbf{0}_{F \times T}\}} \frac{\|\bZ- \cP_{\Phi}(\bZ)\|_F}{\|\bZ\|_F} \nonumber\\
&\stackrel{(e)}{\geq} c^{1/2}(1-\delta_k(\bR))^{1/2} \left( 1- \max_{\bZ\in\Omega_R\backslash\{\mathbf{0}_{L\times
T}\}} \frac{\|\cP_{\Phi}(\bZ)\|_F}{\|\bZ\|_F}  \right) \nonumber\\
&\stackrel{(f)}{=} c^{1/2}(1-\delta_k(\bR))^{1/2}(1-\mu(\Phi,\Omega_R)).\nonumber
\end{align}
In obtaining (c), the assumption $\delta_k(\bR)<1$
along with the fact that no column of $\bH$ has more than $k$ nonzero elements
was used to ensure that $\bR\bH \neq \mathbf{0}_{L \times T}$.
In addition, (d) and (f) follow from the definitions \eqref{eq:rip_ineq} and
\eqref{eq:mau_mat}, respectively, while (e) follows from the triangle inequality.
\hfill$\blacksquare$


\noindent\normalsize \emph{\textbf{C. Proof of Lemma~\ref{lem:lemma_4}}}:
Towards establishing the first bound,
from the submultiplicative property of the spectral norm one obtains
\begin{align}
\|\bQ\|=\|\bA_{\Omegaper}\bA_{\Omega}'\left(\bA_{\Omega}\bA_{\Omega}'\right)^{-1}\|& \leq \|\bA_{\Omegaper}\|
\|\bA_{\Omega}'\left(\bA_{\Omega}\bA_{\Omega}'\right)^{-1}\|. \label{eq:bnd_spec_aomega_1}
\end{align}
Next, upper bounds are derived for both factors on the right-hand side
of \eqref{eq:bnd_spec_aomega_1}. First, using the fact that
$\bA'\bA=\bA_{\Omega}'\bA_{\Omega}+\bA_{\Omegaper}'\bA_{\Omegaper}$ one arrives at
\begin{align}
\|\bA_{\Omegaper}\|^2 &= \max_{\bx \neq \mathbf{0}}\frac{\bx'\bA_{\Omegaper}'\bA_{\Omegaper}\bx}{\|\bx\|^2}
=\max_{\bx \neq \mathbf{0}}\frac{\bx'(\bA'\bA-\bA_{\Omega}'\bA_{\Omega})\bx}{\|\bx\|^2}\nonumber\\
& \leq \max_{\bx \neq \mathbf{0}}\frac{\bx'\bA'\bA\bx}{\|\bx\|^2} - \min_{\bx \neq
\mathbf{0}}\frac{\bx'\bA_{\Omega}'\bA_{\Omega}\bx}{\|\bx\|^2}
= \|\bA\|^2 - \sigma_{\text{min}}^2(\bA_{\Omega}'). \label{eq:bnd_spec_aomega_2}
\end{align}
Note that $\bA_{\Omega}'\left(\bA_{\Omega}\bA_{\Omega}'\right)^{-1}$ is the
pseudo-inverse of the full row rank matrix $\bA_{\Omega}$ (cf. Lemma \ref{lem:lemma_3}), and thus
$\|\bA_{\Omega}'\left(\bA_{\Omega}\bA_{\Omega}'\right)^{-1}\|=\sigma_{\text{min}}^{-1}(\bA_{\Omega}')$~\cite{Horn}.
Substituting these two bounds into \eqref{eq:bnd_spec_aomega_1} yields
\begin{align}
\|\bA_{\Omegaper}\bA_{\Omega}'\left(\bA_{\Omega}\bA_{\Omega}'\right)^{-1}\|& \leq \left\{ \left(\frac{\|\bA\|}{\sigma_{\text{min}}(\bA_{\Omega}')}\right)^2 - 1\right\}^{1/2}.\label{eq:bnd_spec_aomega_3}
\end{align}
In addition, it holds that
\begin{align}
\|\bA\|^2 & = \lambda_{\text{max}}\left\{(\bI-\bP_V) \otimes \bR'(\bI-\bP_U)\bR \right\}\nonumber\\
&=\lambda_{\text{max}}\{(\bI-\bP_V)\} \times \lambda_{\text{max}}\left\{ \bR'(\bI-\bP_U)\bR \right\}\nonumber\\
&\stackrel{(a)}{=}  \|\bR'(\bI-\bP_U)\|^2
\stackrel{(b)}{=} 1. \label{eq:bnd_spec_aomega_4}
\end{align}
where in (a) and (b) it was used that the rows of $\bR$ are orthonormal, and
the maximum singular value of a projection matrix is one.
Substituting~\eqref{eq:bnd_spec_aomega_4} and the bound of Lemma~\ref{lem:lemma_3}
into~\eqref{eq:bnd_spec_aomega_3}, leads to \eqref{eq:bnd_spectnorm_psudo}.

In order to prove the second bound,
first suppose that $\|\bI-\bA_{\Omega}\bA_{\Omega}'\|_{\infty,\infty} < 1$. Then, one can write
\begin{align}
\|\bA_{\Omegaper}\bA_{\Omega}'\left(\bA_{\Omega}\bA_{\Omega}'\right)^{-1}\|_{\infty,\infty} & = \|\bA_{\Omegaper}\bA_{\Omega}'\|_{\infty,\infty} \|\left(\bA_{\Omega}\bA_{\Omega}'\right)^{-1}\|_{\infty,\infty}\nonumber\\
&\leq \|\bA_{\Omegaper}\bA_{\Omega}'\|_{\infty,\infty} \|\left(\bI - (\bI-\bA_{\Omega}\bA_{\Omega}')\right)^{-1}\|_{\infty,\infty}\nonumber\\
&\leq \frac{\|\bA_{\Omegaper}\bA_{\Omega}'\|_{\infty,\infty}}
{1-\|\bI-\bA_{\Omega}\bA_{\Omega}'\|_{\infty,\infty}}.\label{eq:bnd_init}
\end{align}
In what follows, separate upper bounds are derived for
$\|\bA_{\Omegaper}\bA_{\Omega}'\|_{\infty,\infty}$ and $\|\bI-\bA_{\Omega}\bA_{\Omega}'\|_{\infty,\infty}$. For
notational convenience introduce $\cS:=\supp(\bA_0)$ (resp. $\bar{\cS}$ denotes the set complement).
Starting with the numerator in the right-hand side of \eqref{eq:bnd_init}
\begin{align}
\|\bA_{\Omegaper}\bA_{\Omega}'\|_{\infty,\infty} &= \max_{i} \|\be_i'\bA_{\Omegaper}\bA_{\Omega}'\|_1
=\max_{i}\sum_{k}|\langle \be_i'\bA_{\Omegaper},\be_k'\bA_{\Omega} \rangle|\nonumber\\
&=\max_{j}\sum_{\ell}|\langle \be_j'\bA,\be_{\ell}'\bA \rangle|
=\max_{j}\sum_{\ell}|\langle \bA\bA' \be_j,\be_{\ell} \rangle|\nonumber\\
&=\max_{(j_1,j_2)\in \bar{\cS}} \sum_{({\ell}_1,{\ell}_2)\in\cS}|\langle \bR'(\bI-\bP_{U})\bR\be_{j_1}\be_{j_2}'(\bI-\bP_{V}),\be_{\ell_1} \be_{\ell_2}'  \rangle|\nonumber\\
&=\max_{(j_1,j_2) \in \bar{\cS}} \sum_{({\ell}_1,{\ell}_2) \in \cS}\underbrace{|\langle \bR\be_{j_1}
\be_{j_2}'(\bI-\bP_{V}),(\bI-\bP_{U})\bR\be_{\ell_1} \be_{\ell_2}'  \rangle|}_{:=g(j_1,j_2,\ell_1,\ell_2)}.
\label{eq:bnd_term1_1}
\end{align}
Following some manipulations, the term inside the summation can be further bounded as
\begin{align}
g(j_1,j_2,\ell_1,\ell_2) & = |\langle \bR \be_{j_1} \be_{j_2}', (\bI-\bP_U)\bR \be_{\ell_1}\be_{\ell_2}'\rangle - \langle \bR \be_{j_1} \be_{j_2}'\bP_{V},(\bI-\bP_U)\bR \be_{\ell_1} \be_{\ell_2}' \rangle \nonumber\\
&= |\langle  \be_{j_2}'\be_{\ell_2}, \be_{j_1}'\bR'(\bI-\bP_U)\bR\be_{\ell_1} \rangle - \langle  \be_{j_2}'\bP_{V}\be_{\ell_2}, \be_{j_1}'\bR' (\bI-\bP_U)\bR \be_{\ell_1}  \rangle    \nonumber\\
&= | \be_{j_1}'\bR'(\bI-\bP_U)\bR\be_{\ell_1} \ind_{\{j_2=\ell_2\}} - (\be_{j_2}'\bP_{V}\be_{\ell_2}) (\be_{j_1}'\bR'
(\bI-\bP_U)\bR \be_{\ell_1}) |.
\end{align}
Upon defining $x_{j_1,\ell_1}:=\be_{j_1}'\bR'(\bI-\bP_U)\bR\be_{\ell_1}$ and
$y_{j_2,\ell_2}:=(\be_{j_2}'\bP_{V}\be_{\ell_2})$, squaring $g$ gives rise to
\begin{align}
g^2(j_1,j_2,\ell_1,\ell_2) =  x_{j_1,\ell_1}^2 \ind_{\{j_2=\ell_2\}} + y_{j_2,\ell_2}^2 x_{j_1,\ell_1}^2 - 2
y_{j_2,\ell_2} x_{j_1,\ell_1}^2 \ind_{\{j_2=\ell_2\}}.\label{eq:g^2}
\end{align}
Since $y_{j_2,\ell_2}\ind_{\{j_2=\ell_2\}}=\|\bP_V \be_{j_2}\|^2 \ind_{\{j_2=\ell_2\}} \geq 0$, one can ignore the
third summand in \eqref{eq:g^2} to arrive at
\begin{align}
g(j_1,j_2,\ell_1,\ell_2) \leq  x_{j_1,\ell_1} [\ind_{\{j_2=\ell_2\}} + y_{j_2,\ell_2}^2]^{1/2}.  \label{eq:bnd_g^2}
\end{align}
Towards bounding the scalars $x_{j_1,\ell_1}$ and $y_{j_2,\ell_2}$, rewrite
$x_{j_1,\ell_1}:=\be_{j_1}'\bR'\bR\be_{\ell_1}-\be_{j_1}'\bR'\bP_U\bR\be_{\ell_1}$.
If $j_1=\ell_1$, it holds that $x_{j_1,\ell_1} \leq \|\bR\be_{\ell_1}\|^2 \leq c(1+\delta_1(\bR))$; otherwise,
\begin{align}
x_{j_1,\ell_1} \leq  |\be_{j_1}'\bR'\bR\be_{\ell_1}| + |\be_{j_1}'\bR'\bP_U\bR\be_{\ell_1}| \
\leq c\theta_{1,1}(\bR) + c(1+\delta_1(\bR)) \gamma_R^2(\bU). \nonumber
\end{align}
Moreover, $y_{j_2,\ell_2} \leq \|\bP_V\be_{j_2}\| \|\bP_V\be_{\ell_2}\| \leq \gamma^2(\bV)$.
Plugging the bounds into \eqref{eq:bnd_g^2} yields
\begin{align}
g(j_1,j_2,\ell_1,\ell_2) \leq & \left[ c(1+\delta_1(\bR))\ind_{\{j_1=\ell_1\}} + c(\theta_{1,1}(\bR) +
c(1+\delta_1(\bR)) \gamma_R^2(\bU) ) \ind_{\{j_1 \neq \ell_1\}} \right] \nonumber\\
& \times [\ind_{\{j_2=\ell_2\}} + \gamma^4(\bV)]^{1/2}. \label{eq:bnd_g}
\end{align}
Plugging \eqref{eq:bnd_g} into \eqref{eq:bnd_term1_1} one arrives at
\begin{align}
\|\bA_{\Omegaper}\bA_{\Omega}'\|_{\infty,\infty}
 {}\leq{} & c[\sqrt{2}k + s\gamma^2(\bV) ]\theta_{1,1}(\bR) + c(1+\delta_1(\bR)) \left[k \gamma^2(\bV) +
 \sqrt{2}k \gamma_R^2(\bU) + s \gamma_R^2(\bU) \gamma^2(\bV) \right]\nonumber\\
 {}:={}& c \omega_{\text{max}} \label{eq:bnd_term1}
\end{align}
after using: i) $\cS \cap \bar{\cS}=\emptyset$ and consequently
$j_2 \neq \ell_2$ when $j_1=\ell_1$; and ii) $\gamma(\bV) \leq 1$.

Moving on, consider bounding $\|\bI-\bA_{\Omega}\bA_{\Omega}'\|_{\infty,\infty}$ that
can be rewritten as
\begin{align}
\|\bI-\bA_{\Omega}{\bA_{\Omega}}'\|_{\infty,\infty} &= \max_{i} \|{\be_i}'(\bI-\bA_{\Omega}{\bA_{\Omega}}')\|_1 \nonumber\\
&=\max_{i} \left\{ |1 - \|{\be_i}'\bA_{\Omega}\|^2 | + \sum_{k \neq i} | \langle {\be_i}'\bA_{\Omega},{\be_k}'\bA_{\Omega} \rangle | \right\} \nonumber\\
&=\max_{\substack{j=j_1+j_2\\(j_1,j_2) \in \cS}} \left\{|1-\|\bA'\be_j\|^2 | + \sum_{\ell \neq j}|\langle \bA'\be_j,\bA'\be_{\ell} \rangle|\right\}.\label{eq:bound_term2_1}
\end{align}
In the sequel, an upper bound is derived for \eqref{eq:bound_term2_1}. Let $(j_1,j_2)$
denote the element of $\cS$ associated with $j$ in~\eqref{eq:bound_term2_1}.
For the first summand inside the curly brackets in \eqref{eq:bound_term2_1}, consider
lower bounding the norm of the $j$-th row of $\bA$ as
\begin{align}
\|\bA'\be_j\| &= \|(\bI-\bP_{U})\bR\be_{j_1} \be_{j_2}'(\bI-\bP_{V})\|_F
= \|\cP_{\Phiper}(\bR\be_{j_1} \be_{j_2}')\|_F \nonumber\\
& = \|\bR\be_{j_1} \be_{j_2}' - \cP_{\Phi}(\bR\be_{j_1} \be_{j_2}')\|_F
\geq \|\bR\be_{j_1} \be_{j_2}'\| - \| \cP_{\Phi}(\bR\be_{j_1} \be_{j_2}')\|_F \nonumber\\
& \geq \|\bR\be_{j_1} \be_{j_2}'\| (1-\mu(\Phi,\Omega_R))
\geq c^{1/2}(1-\delta_1(\bR))^{1/2} (1-\mu(\Phi,\Omega_R)).\nonumber
\end{align}
Since $\delta_1(\bR) < 1$ and $\mu(\Phi,\Omega_R) < 1$, one obtains
$|1 - \|\bA' \be_j\|^2 | \leq 1-c(1-\delta_1(\bR)) (1-\mu(\Phi,\Omega_R))^2 $.

For the second summand inside the curly brackets in \eqref{eq:bound_term2_1},
a procedure similar to the one used for bounding
$\|\bA_{\Omegaper}{\bA_{\Omega}}'\|_{\infty,\infty}$ is pursued. First, observe that
\begin{align}
 \sum_{\ell \neq j} | \langle \bA\bA'\be_j,\be_{\ell} \rangle |
&=\sum_{\ell \neq j} | \langle (\bI-\bP_{V})\otimes \bR'(\bI-\bP_{U})\bR\be_j,\be_{\ell} \rangle | \nonumber\\
&=\sum_{(\ell_1,\ell_2) \in \cS \backslash \{(j_1,j_2)\} } | \langle \bR'(\bI-\bP_{U}) \bR \be_{j_1} \be_{j_2}'(\bI-\bP_{V}),\be_{\ell_1} \be_{\ell_2}' \rangle | \nonumber\\
&=\sum_{(\ell_1,\ell_2) \in \cS \backslash \{(j_1,j_2)\} }  | \langle \bR \be_{j_1}
\be_{j_2}'(\bI-\bP_{V}),(\bI-\bP_{U}) \bR \be_{\ell_1} \be_{\ell_2}' \rangle |  \label{eq:summation}
\end{align}
to deduce that, up to a summand corresponding to the index pair $(j_1,j_2)$, \eqref{eq:summation} is
identical to the summation in \eqref{eq:bnd_term1_1}.
Following similar arguments to those leading to \eqref{eq:bnd_g}, one arrives at
\begin{equation*}
\max_{\substack{j=j_1+j_2\\(j_1,j_2) \in \cS}}\sum_{\ell \neq j} | \langle \bA'\be_j,\bA'\be_{\ell} \rangle |  \leq c
\omega_{\text{max}}.
\end{equation*}

Putting pieces together, \eqref{eq:bound_term2_1} is bounded as
\begin{align}
\|\bI - \bA_{\Omega}\bA_{\Omega}'\|_{\infty,\infty}  \leq &
1 - c(1-\delta_1(\bR)) (1-\mu(\Phi,\Omega_R))^2 + c \omega_{\text{max}}. \label{eq:bnd_term2}
\end{align}
Note that because of the assumption $\omega_{\text{max}} <
(1-\delta_1(\bR)) (1-\mu(\Phi,\Omega_R))^2$, $\|\bI -
\bA_{\Omega}\bA_{\Omega}'\|_{\infty,\infty} <1$ as supposed at the beginning
of the proof. Substituting~\eqref{eq:bnd_term1} and \eqref{eq:bnd_term2} into \eqref{eq:bnd_init} yields the desired bound.\hfill$\blacksquare$


\noindent\normalsize \emph{\textbf{D. Proof of Lemma~\ref{lem:lemma_5}}}:
The proof bears some resemblance with those available for the matrix
completion problem~\cite{CR08}, and PCP~\cite{CLMW09}.
However, presence of the compression matrix $\bR$ gives rise
to unique challenges in some stages of the proof,
which necessitate special treatment. In what follows, emphasis is placed on
the distinct arguments required by the setting here.

The main idea is to obtain first an upper bound on the norm of the linear operator
$\pi^{-1}\cP_{\Phi}\bR\cP_{\Omega}\bR'\cP_{\Phi}-\cP_{\Phi}$,
which is then utilized to upper bound $\mu(\Phi,\Omega_R)=\|\cP_{\Phi}\bR\cP_{\Omega}\|$.
The former is established in the next
lemma; see Appendix E for a proof.

\begin{lemma} \label{lem:lemma_10}
Suppose $\cS:=\supp(\bA_0)$ is drawn according to the Bernoulli model with parameter $\pi$. Let
$\Lambda:=\sqrt{c(1+\delta_1(\bR)) [\gamma_R^2(\bU) + \gamma^2(\bV)]}$, and $n:=\max\{L,F\}$.
Then, there are positive numerical constants $C$ and $\tau$ such that
\begin{align}
\pi^{-1} \|\cP_{\Phi}\bR\cP_{\Omega}\bR'\cP_{\Phi} - \pi \cP_{\Phi} \| \leq C \sqrt{\frac{\log(LF)}{\pi}} + \tau \Lambda \log(n) \label{eq:bnd_operatornorm}
\end{align}
holds with probability higher than $1-\mathcal{O}\left(n^{-C\pi\Lambda\tau}\right)$, provided
that the right-hand side is less than one.
\end{lemma}

Building on \eqref{eq:bnd_operatornorm}, it follows that
\begin{align}
\|\cP_{\Phi}\bR\cP_{\Omega}\bR'\cP_{\Phi}\| - \pi &\stackrel{(a)}{\leq} \|\cP_{\Phi}\bR\cP_{\Omega}\bR'\cP_{\Phi}\| -
\pi \|\cP_{\Phi}\| \nonumber\\
&\stackrel{(b)}{\leq}  \|\cP_{\Phi}\bR\cP_{\Omega}\bR'\cP_{\Phi} - \pi\cP_{\Phi}\| \nonumber\\
&\leq C \sqrt{\pi\log(LF)} + \tau \pi \Lambda \log(n) \label{eq:bnd_diff}
\end{align}
where (a) and (b) come from $\|\cP_{\Phi}\| \leq 1$ and the triangle inequality, respectively.
In addition,
\begin{align}
\|\cP_{\Omega}(\bR'\cP_{\Phi}(\bX))\|_F^2 &= |\langle \cP_{\Omega}(\bR'\cP_{\Phi}(\bX)), \cP_{\Omega}(\bR'\cP_{\Phi}(\bX))  \rangle| \nonumber\\
& = |\langle \cP_{\Phi}(\bR(\cP_{\Omega}(\bR'\cP_{\Phi}(\bX)))), \bX  \rangle| \nonumber\\
& \leq \|\cP_{\Phi}(\bR(\cP_{\Omega}(\bR'\cP_{\Phi}(\bX))))\|_F \|\bX\|_F \label{eq:bnd_mu_x}
\end{align}
for all $\bX \in \mathbb{R}^{L \times F}$.
Recalling the definition of the operator norm, it follows
from~\eqref{eq:bnd_mu_x} that $\mu(\Phi,\Omega_R) \leq
\sqrt{c^{-1}(1-\delta_k(\bR))^{-1}}\|\cP_{\Phi}\bR\cP_{\Omega}\bR'\cP_{\Phi}\|^{1/2}$.
Plugging the bound~\eqref{eq:bnd_diff}, the result follows readily.\hfill$\blacksquare$


\noindent\normalsize \emph{\textbf{E. Proof of Lemma~\ref{lem:lemma_10}}}:
Start by noting that
\begin{align}
\bR'\cP_{\Phi}(\bX)= &\sum_{i,j} \langle \bR'\cP_{\Phi}(\bX), \be_i\be_j' \rangle \be_i\be_j'
=  \sum_{i,j} \langle \bX, \cP_{\Phi}(\bR \be_i\be_j') \rangle \be_i\be_j' \nonumber 
\end{align}
and apply the sampling operator to obtain
\begin{align}
\cP_{\Omega}(\bR'\cP_{\Phi}(\bX)) = \sum_{i,j} b_{i,j} \langle \bX, \cP_{\Phi}(\bR \be_i\be_j') \rangle \be_i\be_j'
\nonumber 
\end{align}
where $\{b_{i,j}\}$ are Bernoulli-distributed i.i.d. random variables with ${\rm Pr}(b_{i,j}=1)=\pi$.
Then,
\begin{align}
\cP_{\Omega}(\bR \cP_{\Omega}(\bR'\cP_{\Phi}(\bX))) = \sum_{i,j} b_{i,j} \langle \bX, \cP_{\Phi}(\bR \be_i\be_j')
\rangle \cP_{\Phi}(\bR \be_i\be_j').
\end{align}
Moreover, since $\bR\bR'=\bI_L$ one finally arrives at
\begin{align}
\cP_{\Phi}(\bX) = \cP_{\Phi}(\bR\bR'\cP_{\Phi}(\bX)) = \sum_{i,j} b_{i,j} \langle \bX, \cP_{\Phi}(\bR
\be_i\be_j') \rangle \cP_{\Phi}(\bR \be_i\be_j'). \label{eq:P_phi_R_Pomega_R_prime_P_Phi_X}
\end{align}
The next bound will also be useful later on
\begin{align}
\|\cP_{\Phi}(\bR\be_i{\be_j}')\|_F^2 & = \langle \cP_{\Phi}(\bR\be_i{\be_j}'), \bR\be_i{\be_j}'   \rangle \nonumber\\
& = \langle \bP_{U} \bR \be_i {\be_j}' + \bR \be_i {\be_j}' \bP_{V} - \bP_{U}
\bR \be_i {\be_j}' \bP_{V}  , \bR \be_i {\be_j}'  \rangle \nonumber\\
& = \langle \bP_{U} \bR \be_i {\be_j}', \bR\be_i {\be_j}'  \rangle +
\langle \bR \be_i {\be_j}'\bP_{V}, \bR \be_i {\be_j}'  \rangle -
\langle \bP_{U} \bR \be_i {\be_j}' \bP_{V}, \bR \be_i {\be_j}'  \rangle  \nonumber\\
& \stackrel{(a)}{=} \|\bP_{U} \bR \be_i {\be_j}'\|_F^2 + \|\bR \be_i
{\be_j}'\bP_{V}\|_F^2 - \|\bP_{U} \bR \be_i {\be_j}'\|_F^2  \|\bP_{V} \be_j\|_F^2\nonumber\\
& \leq c(1+\delta_1(\bR)) \gamma_R^2(\bU) + c(1+\delta_1(\bR)) \gamma^2(\bV)
=\Lambda^2 \label{eq:bnd_Pphieiej}
\end{align}
where (a) holds because $\langle \bP_{U} \bR \be_i {\be_j}' \bP_{V}, \bR \be_i {\be_j}'
\rangle=\langle \be_i' \bR \bP_{U} \bR \be_i,  \be_j' \bP_{V}\be_j  \rangle$
and $\bP_{U}=\bP_{U}^2$ (likewise $\bP_{V}$).

Defining the random variable
$\Xi:=\pi^{-1} \|\cP_{\Phi}\bR\cP_{\Omega}\bR'\cP_{\Phi} - \pi \cP_{\Phi} \|$ and using \eqref{eq:P_phi_R_Pomega_R_prime_P_Phi_X}, one can write
\begin{align}
\Xi ={}& \pi^{-1} \sup_{\|\bX\|_F=1} \Big\|\sum_{i,j}(b_{i,j}-\pi) \langle \bX,\cP_{\Phi}(\bR \be_i\be_j')  \rangle
\cP_{\Phi}(\bR \be_i\be_j') \Big\|_F \nonumber\\
={}& \pi^{-1} \sup_{\|\vec(\bX)\|=1} \Big\|\sum_{i,j}(b_{i,j}-\pi)  \vec(\bX)' \vec[\cP_{\Phi}(\bR \be_i\be_j')]
\otimes \vec[\cP_{\Phi}(\bR \be_i\be_j')] \Big\| \nonumber \\
={}& \pi^{-1} \Big\|\sum_{i,j}(b_{i,j}-\pi) \vec[\cP_{\Phi}(\bR \be_i\be_j')] \otimes \vec[\cP_{\Phi}(\bR
\be_i\be_j')] \Big\|. \label{eq:Xi}
\end{align}
Random variables $\{b_{i,j}-\pi\}$ are i.i.d. with zero mean, and thus one can
utilize the spectral concentration inequality in \cite[Lemma 3.5]{Rudelson}
to find
\begin{align}
\mathbb{E} [\Xi] &\leq C \sqrt{\frac{\log(LF)}{\pi}} \max_{i,j}{\|\cP_{\Phi}(\bR\be_i\be_j')\|_F}
\stackrel{(b)}{\leq}  C \sqrt{\frac{\log(LF)}{\pi}} \Lambda \label{eq:rudelson_ineq}
\end{align}
for some constant $C>0$, where (b) is due to~\eqref{eq:bnd_Pphieiej}. Now, applying Talagrand's
concentration tail bound~\cite{Talagrand} to the random variable $\Xi$ yields
\begin{align}
{\rm Pr}(|\Xi - \mathbb{E}[\Xi]| \geq t) \leq 3 \exp\left( -\frac{t \log(2)}{K} \pi \min\{1,t\}
\right) \label{eq:talagrand_ineq}
\end{align}
for some constant $K > 0$, where $t:=\tau \Lambda \log(n)$ and $n:=\max\{L,F\}$.
The arguments leading to~\eqref{eq:rudelson_ineq} and \eqref{eq:talagrand_ineq} are similar those
used in~\cite[Theorem 4.2]{CR08} for the matrix completion problem, and
details are omitted here. Putting~\eqref{eq:rudelson_ineq} and~\eqref{eq:talagrand_ineq} together
it is possible to infer
\begin{align}
\Xi &\leq \mathbb{E}[\Xi] + t  \leq   C \sqrt{\frac{\log(LF)}{\pi}} + \tau \Lambda \log(n)
\end{align}
with probability higher than $1-\mathcal{O}(n^{-C \pi \Lambda \tau})$, which completes
the proof of the lemma.\hfill$\blacksquare$

\newpage


\bibliographystyle{IEEEtranS}
\bibliography{IEEEabrv,biblio}

\end{document}